\documentclass[12pt]{article}
\usepackage{epsfig}
\begin{document}
\def\l{{(\lambda)}}
\def\tr{{\rm tr}\, }
\def\Tr{{\rm Tr}\, }
\def\hTr{\hat{\rm T}{\rm r}\, }
\def\be{\begin{eqnarray}}
\def\ee{\end{eqnarray}}
\def\ctt{\chi_{\tau\tau}}
\def\cta{\chi_{\tau a}}
\def\ctb{\chi_{\tau b}}
\def\cab{\chi_{ab}}
\def\cba{\chi_{ba}}
\def\ptt{\phi_{\tau\tau}}
\def\pta{\phi_{\tau a}}
\def\ptb{\phi_{\tau b}}
\def\pab{\phi_{ab}}
\def\lb{\label}\def\appendix{{\newpage\section*{Appendix}}\let\appendix\section%
        {\setcounter{section}{0}
        \gdef\thesection{\Alph{section}}}\section}
\renewcommand{\figurename}{Fig.}
\renewcommand\theequation{\thesection.\arabic{equation}}
\renewcommand{\thefootnote}{\fnsymbol{footnote}}
\hfill{\tt IUB-TH-045}\\\mbox{}
\hfill {\tt hep-th/0405252}\\
\vskip0.3truecm
\begin{center}
\vskip 2truecm {\Large\bf Reconstructing Minkowski Space-Time}
\vskip 1truecm
{\large\bf
Sergey N.~Solodukhin\footnote{
{\tt s.solodukhin@iu-bremen.de}}
}\\
\vskip 0.6truecm
\it{School of Engineering and Science, 
International University Bremen, \\
P.O. Box 750561, Bremen 28759, Germany}
\vskip 1truemm
\end{center}
\vskip 1truecm
\begin{abstract}
\noindent
Minkowski space is a physically important space-time for which 
the finding an adequate holographic description is an urgent problem.
In this paper we develop further the proposal made in   
\cite{dBS} for the description as a duality between Minkowski 
space-time and a Conformal 
Field Theory defined on the boundary of the light-cone. We focus on the 
gravitational aspects of the duality. Specifically, we identify  the 
gravitational holographic data and provide the way  Minkowski space-time
(understood in more general context as a Ricci-flat space)
is reconstructed from the data. In order to avoid the 
complexity of non-linear Einstein equations 
we consider linear perturbations and do the analysis for the perturbations. 
The analysis  proceeds in two steps.
We first  reduce the problem in Minkowski space to
an infinite set of field equations on de Sitter space one dimension
lower. These equations are quite remarkable: they describe massless and
massive gravitons  in de Sitter space. In particular, the partially massless
graviton appears naturally in this reduction. In the second step we solve 
the graviton field equations and identify the holographic boundary data. 
Finally, we consider the asymptotic form of the black hole space-time 
and identify the way the information about the mass of the static 
gravitational configuration is encoded in the holographic data.
\end{abstract}
\vskip 1cm
\newpage

\newpage

\section{Introduction}
\setcounter{equation}0
In this paper we continue the study started in \cite{dBS} of the duality 
between Minkowski space-time and a Conformal Field Theory 
defined on the boundary of the light-cone. That semiclassical gravity may 
know something about quantum field theories was first demonstrated by Brown
and Henneaux in 1986 \cite{Brown} who looked at the algebra of 
gravitational constraints generating the asymptotic symmetries
of three-dimensional anti-de Sitter space-time 
and found that those constraints
form conformal Virasoro algebra with calculable central charge. 
This was the first indication in the physical literature that 
asymptotically anti-de Sitter space
encodes some non-trivial information about conformal symmetry and
quantum anomalies in the space one dimension lower. At approximately same time
mathematicians Fefferman and Graham \cite{FG} were interested in the purely
mathematical problem  of finding possible conformal invariants and discovered that
all such invariants are naturally induced from the ordinary metric
invariants  near conformal boundary of the hyperbolic
space one dimension higher. The relation between conformal symmetry 
and the Einstein spaces with
negative cosmological constant thus was established.
 In their analysis they invented a technical tool of asymptotic expansion, the
now famous ``Fefferman-Graham expansion'',
which later on proved to be very useful in the  physical applications.

For almost a decade the two sides (physical and mathematical) 
of the story followed in parallel without making any close contact.
This was until the holographic principle \cite{tHooft}, \cite{Susskind}
was formulated and a concrete realization of this principle, 
the AdS/CFT correspondence, was suggested \cite{Malda}, \cite{Gubs},
\cite{Wit}.  According to the holographic idea the space-time physics
of gravitationally interacting particles should be more economically described
in terms of some theory living on the boundary (the so-called
``holographic screen''). In the AdS/CFT correspondence the bulk space-time is
anti-de Sitter space and the theory on the boundary is a quantum conformal field
theory. It is possible to formulate  precise bulk/boundary dictionary  
translating the (super)gravity phenomena in the bulk to that of CFT on the 
boundary and vise versa.  This led to many interesting developments. 
Among others, it was understood that there is a deep relation between
geometry of the negative constant curvature space-time and quantum properties
of the conformal theories.  In particular, the conformal anomalies can be
calculated purely geometrically \cite{HS} by first expanding the bulk Einstein 
metric near the conformal boundary  and then inserting the expansion
back to the gravitational action. The Fefferman-Graham expansion thus made its
new appearance and helped to reproduce the Brown-Henneaux central charge
in a purely geometrical fashion. The asymptotic diffeomorphisms in anti-de
Sitter space play an important role generating the conformal symmetry
at the boundary \cite{ISTY} and imposing severe constraints on the 
possible form of the anomalies \cite{Theisen}, \cite{Theisen1}.
Not only conformal anomalies but also the whole structure of the anomalous
stress-tensor of quantum CFT might be possible to extract from the 
geometry of hyperbolic space \cite{KS}, \cite{Rob}. However, with the exception of
the three-dimensional case \cite{KS} this is still an open problem.
An important element in the holographic description is the way how the
hologram should be decoded, i.e. how the bulk gravitational physics is restored from
the boundary CFT data. In \cite{deHaro} the necessary holographic data were
found to be the metric representing the conformal class on the 
boundary and the boundary CFT stress tensor. The ref.\cite{deHaro} then gives 
precise prescriptions for how the space-time metric can be reconstructed from
these data.

The success of the AdS/CFT duality has motivated the attempts to extend the
holographic description to other spaces. It was rather natural 
to generalize it first to de Sitter space, many elements of this new duality 
extend straightforwardly from the anti-de Sitter case while many new
subtleties arise \cite{Wit1}, \cite{Str}, \cite{Jan}, \cite{st2}.
One of them is the problem with unitarity since  typical conformal
weights arising in the duality with de Sitter space are complex. Another is
the problem of formulating the S-matrix description in de Sitter space.
Both problems are still open although some suggestions have been made \cite{Jan1},
\cite{Volovich}.

Minkowski space-time is another important space a holographic description of
which should be understood. A number of ideas and proposals has been circulated
in the literature \cite{Susskind1}-\cite{Krasnov}. In \cite{dBS} it was suggested
to associate the holographic picture with a choice of light-cone in Minkowski
space. The part of the space-time which is out-side the light-cone is
naturally foliated with de Sitter slices while the part which is inside
is sliced with the Euclidean anti-de Sitter spaces. The only boundary of these
slices is the boundary of the light-cone itself which is suggested to be the
place where the holographic data should be collected. Formulating the
holographic dictionary one can make
use of the known prescriptions of the AdS/CFT and dS/CFT dualities applying
these prescriptions to each separate slice and then summing over all slices.
The details of this procedure have been worked out in \cite{dBS}.
In fact, this line of reasoning follows the inspirational  paper \cite{FG}
where the Euclidean hyperbolic space was considered in the context of the 
cone structure in flat space one dimension higher.
The symmetry plays an important role in the identifying the way the
holographic data should be presented. The Lorentz group of
(d+2)-dimensional  Minkowski space becomes the conformal group acting on
d-sphere lying at (past or future) infinity of the light-cone. 
The data thus are expected to have a CFT representation.
In this picture it is natural that the propagating near null infinity
plane waves are dually described by an infinite set of the conformal 
operators living on the d-sphere. Moreover the quantum-mechanical 
S-matrix can be restored in terms of
the correlation functions of operators on two d-spheres: at infinite past and
infinite future on the light-cone.

In the present study we extend the picture suggested in \cite{dBS} and apply
it to the gravitational field itself. More specifically, we want to identify
the minimal set of data which has to be specified at the boundary of the
light-cone and which is sufficient for complete reconstruction of the 
bulk Minkowski space-time.
We understand Minkowski space in a wide sense as a Ricci-flat space-time
asymptotically approaching the standard flat space structure. 
It should be noted that there have been earlier attempts in the literature to proceed  
in a similar direction \cite{BS}, \cite{B}, \cite{deHaro:2000wj}. 
The main idea was to integrate the Einstein equations with zero cosmological
constant starting with the boundary at spatial infinity and developing the series
expansion in the radial direction 
in the similar fashion as Fefferman and Graham have taught us to do in
the case of equations with negative cosmological constant.
This program, however, does not work as nicely for asymptotically flat space
as it did for asymptotically adS space. The recurrent relations between 
terms in the formal series are now differential rather than algebraic as in the
adS case. It is not possible to resolve them and express the 
coefficients in the series in terms of some boundary data at the spatial
infinity. The proposal made in \cite{dBS} is to start at  infinity of the
light-cone and  integrate the equations from there. One has to develop
double expansion in this case: first expansion goes along the
constant-curvature slice 
and the second is in the radial direction enumerating the slices.
The relations between coefficients are now algebraic. They  can be resolved
and the necessary boundary data identified. In order to avoid the 
complexity of  non-linear  gravitational equations we consider the 
linear perturbations and do the analysis for the perturbations. This
certainly simplifies the problem and provides us with the important information on 
the non-linear case as well.

This paper is organized as follows. In section 2 we give some comments on the
holographic reconstruction in general emphasizing the role of the
causality. We use two-dimensional examples to illustrate our point.
In section 3 we review the holographic proposal of \cite{dBS}  
in the case of the scalar field. The way the duality works in
two-dimensional Rindler space is briefly discussed.
We turn to the gravitational case in section 
4 and proceed in two steps. First, we  reduce  the Minkowski problem to
an infinite set of gravitational equations on de Sitter space one dimension
lower. These equations are quite remarkable: they describe massless and
massive gravitons  on de Sitter space. In particular the partially massless
graviton appears naturally in this reduction. In the second step we solve each
graviton field equation on de Sitter space and identify the
boundary data. Decoding the hologram we then have to set the rules and
translate the boundary data to the bulk gravitational physics. 
In section 5 we make a  step in this direction and 
consider the asymptotic form of the 
black hole space-time and identify the holographic data
which encode the information about the mass of the static gravitational 
configuration. We conclude in section 6.

\section{Holographic reconstruction as  a boundary 
value problem}
\setcounter{equation}0
We start with some general comments on the holographic reconstruction
and show that it can be formulated as a (somewhat unusual)
boundary value problem. More specifically it is the problem in which 
the boundary data are entirely
specified on time-like or null-like boundary. To make this discussion
concrete and simple we take a particular example of massless field in 
(1+1)-dimensional space-time. Let's first consider the flat space-time with
coordinates $(t,z)$, the
field equation than takes the form
\be
-\partial_t^2\phi+\partial_z^2\phi=0~~.
\lb{h1}
\ee
Suppose we consider only a part of the space which lies at positive values of 
coordinate $z$. The standard way to formulate the Cauchy problem in this case
would be to specify 1) some initial conditions at 
$t=0$, $\phi(t=0,z)$ and $\partial_t\phi(t=0,z)$; and 2) the boundary
conditions: $\phi (t,z=0)$ or $\partial_z\phi(t,z=0)$. The necessity to have
two pieces of data, one on the ``initial surface'' $t=0$ and another on the 
boundary $z=0$, follows from simple causality argument: in order to determine
the value of the function $\phi(t,z)$ at a point $(t,z)$ we have to have data
inside the past-directed light-cone with the tip at the point $(t,z)$. For
small values of $t$ the light-cone hits only some part of the ``initial
surface'' and the data on that part is sufficient for the determining the
value at the point $(t,z)$. But for large value of $t$ the light-cone starts to
hit the boundary at $z=0$ and the additional data should be specified there.
Thus, two pieces of the data come out very naturally in this standard
formulation.

In the holographic formulation we would like to have only one piece of data,
namely data to be fixed on the time-like boundary $z=0$, and determine the field
$\phi(t,z)$ for all values of $-\infty<t<+\infty$ and $z>0$  from just this data.
Simple causality picture considered above in the case of standard formulation
helps to visualize the problem in this new formulation. We should again draw
a light-cone with the tip at the point $(t,z)$ but now directed towards the
boundary.  The boundary data specified on the part of the boundary 
which lies inside of this light-cone should be sufficient for determining the 
field $\phi$ at the point $(t,z)$. Another words, in order to reconstruct the
field at the point $(t,z)$ from the data on the boundary ($z=0$) we should
be able to communicate with that point by sending  a signal and wait long enough 
to get the signal back from the point $(t,z)$. That's  qualitatively how this
sort of reconstruction should work. Of course, in order to reconstruct  the
field for all points $(t,z)$ we should have at our disposal the all-time boundary data,
i.e. for $-\infty<t<+\infty$. 

As for the question what kind of boundary data should be specified  
the case of  equation (\ref{h1}) shows us that we have to 
specify  a pair of functions
\be
\phi(t,z=0)=\psi(t)~~{\tt and}~~\partial_z \phi(t,z=0)=\chi(t)
\lb{h2}
\ee
on the boundary $z=0$. This pair forms the holographic data on the boundary.
This is typical situation when the holographic data comes in a pair (function
itself and its normal derivative) and we
will call this a ``holographic pair''. The field equation
(\ref{h1}) subject to the boundary conditions (\ref{h2}) can be easily solved
and the solution reads
\be
\phi(t,z)={1\over 2}[\psi(t+z)+\psi(t-z)]+{1\over 2}\int_{t-z}^{t+z}dt'\
\chi(t')~~.
\lb{h3}
\ee
Thus in order to reconstruct the field at a point $(t,z)$ we have to know
boundary value $\phi(t,z=0)$ of the field
in two moments of time: $t+z$ and $t-z$, and the normal
derivative $\partial_z\phi(t,z=0)$ on the boundary 
for all moments of time in-between.

We of course could try to find the solution  as an  expansion in distance from
the boundary. The solution then would take the form of sum 
of two infinite series
\be
\phi(t,z)=\sum_{n=0}^\infty {z^{2n}\over (2n)!}\psi^{(2n)}_t(t)+
\sum_{n=0}^\infty {z^{2n+1}\over (2n)!}\chi^{(2n)}_t(t)~~,
\lb{h4}
\ee
where within each series all coefficients are determined by the boundary 
function $\psi(t)$, $\psi^{(2n)}_t(t)=\partial_t^{(2n)}\psi(t)$
and the function $\chi(t)$, $\chi^{(2n+1)}_t(t)=\partial_t^{(2n+1)}\chi(t)$.
Notice that the causal structure obvious in the complete solution 
(\ref{h3}) is now
invisible in the  expansion (\ref{h4}).

The expansion similar to (\ref{h4}) is the standard way to solve the
holographic boundary value problem for a field in anti-de Sitter space.
The boundary data  then are associated with some   CFT data  
on the boundary. Let's for simplicity consider the two-dimensional anti-de
Sitter space with metric
\be
&&ds^2={d\rho^2\over 4\rho^2}-{1\over \rho}(1-{\rho\over 4})^2dt^2 \nonumber \\
&&=g(\rho)[-dt^2+dz^2]~~,
\lb{h5}
\ee
where we introduced coordinate 
$$
z(\rho)=\ln({2+\sqrt{\rho}\over 2-\sqrt{\rho}})~~.
$$
The anti-de Sitter space has time-like boundary  located at $\rho=0$  
or at $z=0$ in terms of the coordinate $z$. 
The holographic boundary data thus should be specified
there. The metric (\ref{h5}) is conformal to two-dimensional flat space-time,
actually to a part of it with $z\geq 0$.
The massless scalar field
in two dimensions is conformally invariant so that the solution to the
boundary value problem again takes the form (\ref{h3}) where $z$ should be
replaced with $z(\rho)$. The small $\rho$ expansion then has two sort of terms
\be
\phi(t,\rho)=[\psi(t)+\sum_{k=1}^\infty
F_k(t)\rho^k]+\rho^{1/2}[\chi(t)+\sum_{k=1}^\infty
G_k(t)\rho^k]~~,
\lb{h6}
\ee
where $\psi(t)$ is the boundary value of the field $\phi$ and $\chi(t)$ is the
normal derivative of the field at the boundary of anti-de Sitter.
The coefficients $F_k(t)$ are completely determined by $\psi(t)$ and its
derivatives while $G_k(t)$ are determined by $\chi(t)$. Notice again that the
causal structure present in solution (\ref{h3}) (with $z=z(\rho)$) 
is lost when we re-write it in the form of the $\rho$-expansion. In the
adS/CFT correspondence the holographic pair $(\psi(t),\chi(t))$ has the
following interpretation: $\psi(t)$ is associated with the source which
couples to a ``dual'' operator ${\cal O}(t)$ while $\chi(t)$ 
should be associated with
the quantum expectation value of that operator, $\chi(t)= <{\cal O}(t)>$.
The correlation function of the operators at different moments of time can 
be derived according to the standard adS/CFT prescription 
by taking the  normal derivatives of the Green's function
\be
D=-{1\over 2\pi}\ln\tanh{\sigma\over 2}~~,
\lb{green}
\ee
where $\sigma$ is the geodesic distance between two points 
on two-dimensional anti-de Sitter
space. The 2-point function on the boundary of adS$_2$ then reads 
\cite{Solodukhin}
\be
<{\cal O}(t){\cal O}(t')>\sim {1\over \sinh^2({t-t'\over 2})}~~.
\lb{2pt}
\ee
The two-dimensional case is of course too simplistic for applying the adS/CFT
dictionary in full since the ``boundary'' in this case is just a 
point cross the time.  However, this is a good illustration since in higher
dimensions the logic in identifying the holographic data and recovering the
way of reconstructing the bulk physics from that data is essentially the same.
In particular the $\rho$-expansion similar to (\ref{h6}) is the usual tool for
analyzing the reconstruction of supergravity in the bulk from the CFT data
on the boundary of anti-de Sitter. In the case of the gravitational field
itself it is actually the only tool available due to extreme non-linearity of
the gravitational field equations \cite{HS}, \cite{deHaro}.
We, however, want to emphasize the role of causality in the holographic 
reconstruction. This role is not obvious when the local series expansion is used.

It is interesting that the boundary should not be necessarily time-like.
The holographic boundary problem can be set for  a null-like 
boundary. To illustrate this let's once again exploit our two-dimensional example
and consider arbitrary two-dimensional metric which always can be brought to
a conformally flat form
\be
ds^2=e^{\sigma(x_+,x_-)}dx_+dx_-~~.
\lb{h7}
\ee
Let's consider the part of the space-time 
which lies in the conner $x_-\geq 0,\ x_+\leq 0$, the boundary thus consists of
two ``null surfaces'' $x_+=0$ (${\cal H}_+$) and $x_-=0$ (${\cal H}_+$). 
As the boundary data we specify the value of the field
function
\be
\phi(x_+,x_-)|_{{\cal H}_-}=\psi(x_+) ~~{\tt and }~~\phi(x_+,x_-)|_{{\cal H}_+}=\chi(x_-)
 \lb{h8}
\ee
on these null-surfaces.
Since on the intersection of ${\cal H}_+$ and ${\cal H}_-$
the data should agree we have a constraint, $\psi(x_+=0)=\chi(x_-=0)=\phi_H$.
The solution of such formulated boundary value problem  for 
the massless scalar field equation then takes a very simple  form
\be
\phi(x_+,x_-)=\psi(x_+)+\chi(x_-)-\phi_H~~.
\lb{h9}
\ee
Thus, in order to reconstruct the field at the point $(u,v)$ in the bulk we
have to know the boundary data at three points on the null boundary: at point
$x_+=u$ on ${\cal H}_+$, at point $x_-=v$ on ${\cal H}_-$ and at the
bifurcation point $H$ ($x_+=x_-=0$). 

A natural example of the null-surface is horizon in black hole space-time 
or de Sitter space. That horizon can play the role of the holographic screen
and there might be a dual CFT living on the horizon with bulk/boundary
dictionary similar to the one in the case of adS/CFT correspondence was proposed
in \cite{IS}. We refer the reader to that paper for further details.
Another example when the null-surfaces are natural holographic screens
is the Minkowski space-time and the null-screens are the light-cone and
null-infinity. This is a possible way of looking at the Minkowski/CFT duality 
suggested in \cite{dBS}. We discuss this briefly in the next section.

\section{Holographic description in Minkowski space}
\setcounter{equation}0
The holographic construction suggested in \cite{dBS}
is associated with a choice of light-cone.
The null-surface of a given light-cone $\cal C$ naturally splits 
Minkowski spacetime ${\cal M}_{d+2}$ on two regions: the region $\cal A$
lying inside light-cone $\cal C$ and the region $\cal D$ outside
light-cone. The inside region $\cal A$ on the other hand  splits on the
part which is inside the future light-cone (${\cal A}_+$) and the part which is inside the
past light-cone (${\cal A}_-$). Each region admits natural slicing with constant curvature
hypersurfaces. Outside the light-cone it is the slicing with 
$(d+1)$-dimensional de Sitter spaces (which is positive constant curvature
spacetime with Lorentz signature) while inside the light-cone one may choose
the foliation  with Euclidean anti-de Sitter hypersurfaces defined as positive constant 
curvature space with Euclidean signature.  Enumerating the slices we choose
the radial coordinate $r$ in the region $\cal D$ and the time-like coordinate
$t$ in the region $\cal A$. The Minkowski metric then reads
\be
{\cal D}\, :\,\, ds^2=dr^2+r^2(-d\tau^2+\cosh^2\tau\, d\omega^2(\theta)) \nonumber \\
{\cal A}\, :\,\, ds^2=-dt^2+t^2(dy^2+\sinh^2 y\,d\omega^2(\theta))~~,
\lb{1}
\ee
where $(\tau, \theta)$ and $(y,\theta)$ are the coordinates on a de Sitter 
and anti-de Sitter slice respectively, $d\omega^2(\theta)$ is metric on
unit radius d-sphere with angle coordinates $\theta$.




\begin{figure}

\centering 

\epsfig{figure=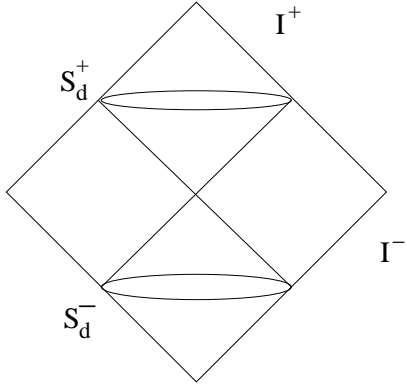, height=2.0in}

\caption{Minkowski space-time with a choice of the light-cone. The boundary sphere
  $S^+_d$ ($S^-_d$) lies in the intersection of the light-cone and
  the future (past) null-infinity.}

\label{fig1}

\end{figure}




%

Each slice in this foliation of Minkowski spacetime is a $(d+1)$-dimensional 
space which has some boundaries. In the anti-de Sitter case the boundary is
the $d$-dimensional sphere $S_d$ lying at infinity of the space while
boundaries of de Sitter space are two spheres $S^+_d$ and $S^-_d$ lying
respectively in the future and in the past of de Sitter space. The considered
foliation has a nice property that all slices have same boundaries as the
light-cone, namely either $S^+_d$ or $S^-_d$.
More precisely, all anti-de Sitter slices covering region ${\cal A}_-$
inside the light-cone have $S^-_d$ as a boundary while $S^+_d$ is the only 
boundary of slices covering region ${\cal A}_+$. Outside the light-cone,
in the region $\cal D$, all de Sitter slices have same boundary $S^+_d$
in the future and $S^-_d$ in the past.
This property motivated the suggestion made in \cite{dBS} to associate the
holographic information on Minkowski space with these two $d$-dimensional spheres.
The Lorentz group of $(d+2)$-dimensional Minkowski space acts as a conformal
group on the spheres $S^+_d$ or $S^-_d$. The holographic information thus is
expected to  have a conformal field theory description. Some of the details of this
holographic description have been demonstrated in \cite{dBS}. An important element of the
description is the specification of  necessary information to be stored 
on the holographic screens as well as the way how the hologram should be
decoded, i.e. the rules of reconstruction the bulk Minkowski physics
from the holographic data on the spheres.

This can be illustrated on the example of a massless scalar field\footnote{The
  consideration is naturally generalized for massive fields and higher spin field equations. The
  case of Dirac fermions was considered in \cite{Loran}.}
described by the field equation $\nabla^2\phi=0$. 
Let's take for concreteness the region 
$\cal D$ outside the light-cone. Then the field equation reads
\be
&&\left(\partial_r^2+{(d+1)\over r}\partial_r+{1\over r^2}\nabla^2_{\tt
    dS}\right) \phi(r,\tau,\theta)=0\, ,~~\nonumber \\
&&\nabla^2_{\tt dS}=-\partial^2_\tau-d\tanh\tau\partial_\tau+\cosh^{-2}\tau\,
\Delta_\theta \, ,
\lb{2}
\ee
where $\Delta_\theta$ is the Laplace operator on unit radius d-sphere. 
Solving equation (\ref{2}) in the region $\cal D$ we expand the field 
$\phi (r,\tau,\theta)$ in powers
of radial coordinate $r$ so that the solution takes the form of Mellin transform
\be
\phi(r,\tau,\theta)={1\over 2\pi i}\int_{{d\over 2}-i\infty}^{{d\over 2}+i\infty} 
d\lambda \, r^{-\lambda}\phi_\lambda (\tau,\theta)~~,
\lb{3}
\ee
where the functions $\phi_\lambda (\tau,\theta)$ satisfy the massive wave
equation on $(d+1)$-dimensional de Sitter space,
\be
(\nabla^2_{\tt dS} -m^2_\lambda)\phi_\lambda(\tau,\theta)=0~,~~
m^2_\lambda=\lambda (d-\lambda))~~.
\lb{4}
\ee
The important question is the range for the spectral parameter $\lambda$. 
A typical configuration in the field theory is a plane wave
for which the relevant spectral parameter is complex, $\lambda={d\over
  2}+i\alpha$, with $\alpha$ changing from minus to plus infinity. This
explains the choice of the limits in the integral (\ref{3}). Later on, the
parameter $\lambda$ is identified with the conformal weight of the dual operator.
Mass term in (\ref{4}) is real and positive, $m^2={d^2\over 4}+\alpha^2$, in
  this case.
Notice that in general the field $\phi(\tau, r,\theta)$ can be a
  superposition of propagating modes as well as solitonic  configurations.
The Coulomb-like potential would be an example of a configuration 
which requires inclusion in (\ref{3}) of terms
with real values of $\lambda$. Indeed, in $d+2$ space-time dimensions the 
Coulomb-like configuration
\be
\phi={Q\over r^{d-1}(\cosh\tau)^{d-1}}
\lb{Col}
\ee
corresponds to $\lambda=d-1$.
We will see that similarly we have to include both complex and real
$\lambda$    in the gravitational case in order to describe  both the 
gravitational waves  and black holes. For description of the latter
the purely real $\lambda$ are appropriate. Notice also that in the case
$\lambda={d\over 2}$ the two independent solutions to the radial differential 
equation (\ref{2}) are $r^{-d/2}$ and $r^{-d/2}\ln r$.

The representation (\ref{3}) is the first step in the holographic reduction:
it reduces the quantum field in Minkowski space to a (infinite) set 
of massive fields living on de Sitter space  of one dimension lower.
As we have discussed in the beginning of this section each de Sitter slice
has two boundaries, $S^+_d$ and $S^-_d$. The next step thus would be to relate 
each solution of the equation (\ref{4}) to the boundary values of the 
functions $\phi_\lambda (\tau,\theta)$. Since on the de Sitter space the
boundary value problem is in fact the initial value problem the 
boundary data should be specified on the surface $S^-_d$.
The general solution of eq.(\ref{4}) is expressed in terms of $P$- and $Q$-Legendre
functions  (see Appendix B)
\be
&&\phi_\lambda(\tau,\theta)=(\cosh\tau)^{1-d\over
  2}[A(\Delta)P_{{(d-1)\over 2}-\lambda}^{\sqrt{(d-1)^2-4\Delta}\over
      2}(-i\sinh\tau){\cal O}_\lambda^>(\theta) \nonumber \\
&&+B(\Delta)Q_{{(d-1)\over 2}-\lambda}^{\sqrt{(d-1)^2-4\Delta}\over
      2}(-i\sinh\tau) {\cal O}^<_\lambda(\theta)]~~,
\lb{phi}
\ee
where the ``constants'' $A(\Delta)$ and $B(\Delta)$ can be chosen in a way to
remove the non-locality in the leading term when expression (\ref{phi}) is 
expanded in powers of $e^\tau$.

The expansion  in powers of $e^{\tau}$ could be used as an alternative way to
solve the equation (\ref{4}).
The coefficients in front of terms
$e^{\lambda\tau}$ and $e^{(d-\lambda)\tau}$ in this expansion are not
determined from the equation: these are the initial data to be specified on $S^-_d$.  
In the form (\ref{phi}) of the solution these are the functions ${\cal
  O}^>_\lambda (\theta)$ and ${\cal O}^<_\lambda (\theta)$ which form the
``boundary'' data.
Combining this expansion with the integral 
representation (\ref{3}) we find that to the leading order near the sphere
$S^-_d$ the solution to
the field equation (\ref{2}) in Minkowski space reads
\be
\phi(r,\tau,\theta)={1\over 2\pi i}\int_{{d\over 2}-i\infty}^{{d\over 2}+i\infty} 
d\lambda (r^{-\lambda} e^{(d-\lambda)\tau} \left[{\cal O}^<_\lambda(\theta)
+\sum_{n=1}^\infty \varphi^<_{(n)\lambda}(\theta)e^{2n\tau}\right]\nonumber\\
+r^{-\lambda} e^{\lambda\tau} \left[{\cal O}^>_\lambda (\theta)+
\sum_{n=1}^\infty \varphi^>_{(n)\lambda}(\theta)e^{2n\tau}\right])~~,
\lb{5}
\ee
where  the higher order terms in the expansion 
are uniquely determined by ${\cal O}^<_\lambda (\theta)$ and ${\cal O}^>_\lambda
(\theta)$. 
The coefficients ${\cal O}^<_\lambda (\theta)$ and ${\cal O}^>_\lambda
(\theta)$ are 
thus the holographic data which are needed to be specified at the sphere
$S^-_d$ for the reconstruction of the scalar field everywhere in the region ${\cal D}$.
These functions can be also associated with the left- and right-moving waves.
The sphere $S^-_d$ lies in the intersection of two null-hypersurfaces:
the past null infinity ${\cal I}^-$ and the past light-cone ${\cal C}_-$. 
Therefore, instead of  two
infinite sets of functions on the d-sphere we may consider just two functions defined on 
null-space of one dimension higher,
\be
{\cal O}^>(\xi,\theta)={1\over 2\pi i}\int_{{d\over 2}-i\infty}^{{d\over 2}+i\infty} 
d\lambda \, \xi^{-\lambda} \, {\cal O}^>_\lambda (\theta) \nonumber \\
{\cal O}^<(\eta,\theta)={1\over 2\pi i}\int_{{d\over 2}-i\infty}^{{d\over 2}+i\infty} 
d\lambda \, \eta^{-\lambda} \, {\cal O}^<_\lambda (\theta)~~,
\lb{6'}
\ee
where $\xi$ ($\eta$) is the affine parameter along the null infinity ${\cal I}^-$
(${\cal C}_-$).

Similar analysis can be done for the solution near the sphere $S^+_d$
in the future of the de Sitter slices. A similar  infinite set
of functions could be specified there. In the quantum mechanical picture
the data on the future sphere $S^+_d$ can be associated with the quantum
out-state while the data specified on $S^-_d$ form the quantum in-state.
In the CFT/Minkowski duality proposed in \cite{dBS} each  coefficient
${\cal O}^{>(<)}_\lambda(\theta)$ appearing
in the expansion (\ref{5}) (and in analogous expansion near $S^+_d$)
is associated with quantum conformal operator of conformal dimension
$\lambda$. The correlation functions of the in- and out-operators are 
\be
&&<0| {_{\tt out}}\!{\cal O}^{<(>)}_{\lambda_1} (\theta_1)~
{_{\tt in}}\!{\cal O}^{>(<)}_{\lambda_2} (\theta_2)|0> \sim ~\delta
(\lambda_1+\lambda_2-d)~\delta^{(d)}(\theta_1,\pi-\theta_2) \nonumber \\
&&<0| {_{\tt out}}\!{\cal O}^{>(<)}_{\lambda_1} (\theta_1)~
{_{\tt in}}\!{\cal O}^{>(<)}_{\lambda_2} (\theta_2)|0> \sim ~\delta
(\lambda_1+\lambda_2-d)
{1\over (1+\cos\gamma(\theta_1,\theta_2))^{\lambda_2}}~~,
\lb{7}
\ee
where $\gamma(\theta,\theta')$ is the geodesic distance
between two points on d-sphere.
The S-matrix of (interacting in the bulk) field than 
can be reconstructed in terms of the correlation functions between 
the conformal operators living on $S^+_d$ and $S^-_d$, as was shown
in \cite{dBS}.

Things are slightly different inside the light-cone, for instance in the
region ${\cal A}_-$ inside the past light-cone. One needs to specify only one
set of functions, namely ${\cal O}_\lambda(\theta)$ there. Or,
equivalently, only a single function on the past null-infinity (more precisely, on
that component of ${\cal I}^-$ which is inside the past light-cone) 
should be specified. This is quite obvious since ${\cal I}^-$ forms a Cauchy
surface in the region ${\cal A}^-$.  In this region the solution to the Cauchy
problem  can be brought to a nice integral form \cite{dBS}
\be
&&\phi(t,y,\theta)={1\over 2\pi i}\int_{{d\over 2}-i\infty}^{{d\over 2}+i\infty} 
d\lambda  \, (-t)^{-\lambda}\int_{S^-_d}d\mu(\theta')G_\lambda(y,\theta,\theta')
{\cal O}_\lambda (\theta') \nonumber \\
&&G_\lambda(y,\theta,\theta')={g(\lambda)\over [\cosh y-\sinh
  y\cos\gamma(\theta,\theta')]^\lambda}~~,
\lb{7'}
\ee
where $g(\lambda)$ is some normalization factor, $d\mu(\theta)$ is the
measure on d-sphere. $G_\lambda(y,\theta,\theta')$ has the meaning of
boundary-to-bulk propagator on $AdS_{d+1}$.

\bigskip

\noindent{\bf Example: (1+1)-dimensional Minkowski space-time}. As an illustration 
we consider a simple example of two-dimensional
Minkowski space with metric
\be
ds^2=-dX_0^2+dX^2_1~~.
\lb{M1}
\ee
The light-cone is defined as $X_0^2-X^2_1=0$, the boundary ``spheres''
$S^+$ and $S^-$ are now  null-dimensional. In two dimensions the region
out-side the light-cone has two components: ${\cal D}_L$ and and ${\cal D}_R$
that are analogous to the region $\cal D$ in the  case of
higher-dimensional Minkowski space-time.
The region out-side the light-cone can be foliated
with hyperbolic curves which are one-dimensional analog of the de Sitter
space-time. The foliation is most transparent in new coordinates $(\tau,r)$ 
defined as
\be
X_0=r\sinh\tau~,~~X_1=\pm r\cosh\tau~~,
\lb{M2}
\ee
where $+$ stands for region ${\cal D}_R$ and $-$ for region ${\cal D}_L$.
In this coordinates the metric takes the form
\be
ds^2=-r^2d\tau^2+dr^2
\lb{M3}
\ee
which can be recognized as two-dimensional Rindler metric. Thus in two
dimensions our general holographic construction naturally leads to Rindler
space. The solution to the field equation takes the form (\ref{5}) with $d=0$
inserted and  all $\varphi_{(n)\lambda}$ vanishing.
In the present case there are  two sets of  operators living on 
the boundary $S^+$ or $S^-$ of the light-cone. These operators, ${\cal O}^>(\omega)$
and ${\cal O}^<(\omega)$, are associated
with left- and right-moving modes there. Notice that there is no angle
dependence since $S^+$ and $S^-$ are just points.
Alternatively, we could consider operators ${\cal O}^>(\xi)$ and 
${\cal  O}^<(\eta)$ living on the past null-infinity and the light-cone and
defined in (\ref{6'}). In this two-dimensional case the reconstruction of the
bulk field in terms of the data on these null-surfaces is especially simple
and is given by expression similar to (\ref{h9}).
The correlation functions of the dual operators can be
read off from the structure of the two-function on the Minkowski space
when each of the functions is approaching one of the boundaries.
Let's restrict our consideration to a simple case of massless field. The
Green's function in this case takes the form
\be
D(X,X')={1\over 4\pi}\ln s^2(X,X')~~,
\lb{M4}
\ee
where the interval $s^2(X,X')$ between two points in terms of the coordinates
$(\tau, r)$ takes the form
\be
s^2=r^2+r'^2-2rr'\cosh(\tau-\tau')~~.
\lb{M5}
\ee
The propagator (\ref{M4}) has a nice representation in terms of the
coordinates $(\tau,r)$. In order to get it we first note that the following
representation of the logarithmic function \cite{HTF}
\be
\ln(1-2zx+x^2)=-1-2\sum_{n=1}^\infty {1\over n}T_n(z)x^n~~
\lb{M6}
\ee
in terms of the Tchebycheff polynomials $T_n(z)=\cos(n\arccos z)$. 
Using this representation
and replacing the infinite sum in (\ref{M6}) with an integral we arrive at another
representation for the propagator (\ref{M4})
\be
D={1\over 4\pi}\ln r^2-{1\over 4\pi}-{1\over 4\pi i}
\int_{\epsilon-i\infty}^{\epsilon+i\infty}{d\lambda\over
  \lambda\sin\pi\lambda}
\cosh\lambda(\tau-\tau') ({r'\over r})^\lambda ~~,
\lb{M7}
\ee
where we have introduced small $\epsilon$ in order to avoid the point
$\lambda=0$ in the integral (\ref{M7}). This propagator is in fact a
superposition of left- and right-moving modes that is easily seen after
the substitution $\lambda=i\omega$ into (\ref{M7})
\be
D={1\over 2\pi}\ln r-{1\over 4\pi}
-{1\over 8\pi}\int_{-\infty-i\epsilon}^{+\infty-i\epsilon} d\omega K(\omega) 
\left(e^{i\omega (\tau-\tau')}({r'\over r})^{i\omega}+e^{-i\omega
    (\tau-\tau')}
({r'\over r})^{i\omega}\right)~~,
\lb{M8}
\ee
where 
$$
K(\omega)={2\over \omega}{1\over e^{2\pi\omega}-1}~~.
$$
From this we find that the correlation function between operators ${\cal O}^>$
and ${\cal O}^<$ reads
\be
<{\cal O}^>(\omega){\cal O}^<(\omega')>={2\over \omega}{1\over
  e^{2\pi\omega}-1}
\delta(\omega+\omega')~~.
\lb{M9}
\ee
Similar expression is valid for ``$>>$'' and ``$<<$'' correlation functions. 
It is interesting to note that the function staying in the right hand side 
in (\ref{M9}) is exactly the thermal factor for the Hawking radiation. The
latter is expected to appear in the Rindler space, the Rindler horizon should
be in fact identified with the light-cone we have chosen. It is quite remarkable
that the conformal operators living on the boundaries of the horizon 
(or the light-cone)  carry certain information about the Hawking radiation.
This observation is another piece of evidence in favor of the so-called 
``horizon holography'' suggested in \cite{IS}. We however do not discuss this
in the present paper.

\section{Reconstruction of metric: linear perturbation \\
analysis}
\setcounter{equation}0
The holographic construction being applied to the metric of space-time
is a somewhat more complicated due to the  dual nature of  the space-time
metric. It   defines the dynamics
of the gravitational field and also sets the background  for 
other fields. Therefore we should consider a class of metrics which
in some sense generalize Minkowski space-time. An appropriate class 
is that of Ricci flat metrics. The gravitational equations thus take the form
\be
R_{\mu\nu}=0~~.
\lb{8}
\ee
These equations are essentially non-linear that makes the analysis more
difficult. Additionally to the field equations (\ref{8}) one has to specify
the boundary conditions, i.e. the asymptotic conditions which should be
satisfied by the metric for describing space-time asymptotically approaching
the ordinary Minkowski space. The meaning of the words ``asymptotically
approaching'' should be specified as well. A natural (from the perspective of the
holographic picture associated with the boundary of the light-cone)
condition formulated in  \cite{dBS} is 
that the space-time should approach the
Minkowski  structure at least in a vicinity of a d-sphere lying in 
the intersection of the null-infinity and the light-cone.
A possible way to solve  the gravitational equations (\ref{8}) subject to this
asymptotic condition is to expand the metric in powers of distance from
the sphere (somewhat analogously to what has been done in the case of
asymptotically anti-de Sitter space). In fact, in the present case a 
double expansion is needed: in powers of $1/r$ of the inverse radial coordinate
enumerating the de Sitter slices close to the sphere $S^-_d$ and
in powers of $e^{\tau}$ measuring the distance to the sphere along the de Sitter slice.
In general this analysis is very complicated. To the leading order in $r$
we however know that the solution to equation (\ref{8}) takes the form \cite{FG}
\be
&&ds^2=dr^2+r^2\,\left(g_{ij}(\tau,\theta)+O(1/r)\right)dx^i\,dx^j~~,\nonumber \\
&&R_{ij}[g]=d\, g_{ij}(\tau,\theta)~~,
\lb{9}
\ee
where $\{x^i\}=\{\tau,\theta\}$. So that to the leading order in $r$
the Ricci flat space-time is foliated with constant positive 
curvature slices generalizing the structure (\ref{1}) 
of the ordinary Minkowski space-time. This asymptotic structure is further
modified by $1/r$ corrections necessarily present in (\ref{9}).
The large $r$ expansion however leads to differential relations between
coefficients in the expansion \cite{BS}, \cite{B}, \cite{deHaro:2000wj}.
These relations can not be solved in general.
On the other hand, expanding  the metric $g_{ij}(\tau,\theta)$ 
in powers of $e^\tau$
we would get the standard asymptotic expansion 
near the boundary of Einstein space of positive constant curvature.
This expansion is algebraic and is similar to the well-known expansion 
for the hyperbolic Einstein space. In most cases this expansion is an infinite
series in $e^{\tau}$. However, when the slice is 3-dimensional the expansion
contains a finite number of terms as was shown in \cite{KS}. 
The boundary $S^-_d$ of the slice is two-dimensional and the conformal
symmetry is infinite-dimensional in this case.
Quite remarkably,
it is the case when the asymptotically flat space-time is 4-dimensional.
This might be an argument for looking more carefully at the physically
interesting 4-dimensional case. 

In this paper we take a different route. Instead of dealing with the
nonlinearity of the equations (\ref{8}) 
we look at linear perturbations of (\ref{8}) around Minkowski space
with metric (\ref{1}).  The holographic analysis then boils down to
applying the holographic construction reviewed in the previous section
to the equations giving the linear perturbations of (\ref{8}).
This is certainly a much simpler problem than solving the non-linear
equations but it also tells us on the possible structure of the 
solution to the non-linear problem and on the necessary holographic
data to be specified for the equation (\ref{8}).

Starting the linear perturbation analysis we re-write the Minkowski metric
(\ref{1}) in the form
\be
ds^2=dr^2+r^2\, \gamma_{ij}(x)dx^i\,dx^j\,\, ,
\lb{11'}
\ee
where $\gamma_{ij}(x)$ is metric on $(d+1)$-dimensional de Sitter space,
\be
&&\gamma_{ij}(x)dx^i\,dx^j=-d\tau^2+e^{A(\tau)}\,\beta_{ab}(\theta)\,
d\theta^a\, d\theta^b\,\, ,
\nonumber \\
&&A(\tau)=2\ln\cosh\tau~~,
\lb{11}
\ee
and $\beta_{ab}(\theta)$ is the metric on unite radius d-sphere.

The equation for linear perturbations $h_{\mu\nu}(r,x)$ takes the form
\be
\nabla^\alpha\nabla_\mu h_{\nu\alpha}+\nabla^\alpha\nabla_\nu h_{\mu\alpha}
-\nabla^\alpha\nabla_\alpha h_{\mu\nu}-\nabla_{\mu}\nabla_\nu {\hTr}h=0~~,
\lb{12}
\ee
where ${\hTr}h$ and covariant derivative $\nabla_\mu$ are defined 
with respect to the Minkowski
metric (\ref{11'}). As in the case of scalar field we further proceed in two
steps.

\subsection{First step: reduction to field equations on de Sitter slice}
First we fix the gauge, $h_{rr}=h_{ri}=0$ so that the only non-vanishing
components are $h_{ij}(r,x)$. In particular, we have that 
$\hTr h=r^{-2}\Tr  h$, $\Tr h=\gamma^{ij}h_{ij}$. The equations (\ref{12}) 
then reduce to a set of equations
\be
   r^2\Tr h''-2r\Tr h'+2\Tr h=0
\lb{13}
\ee
\be
   r\partial_r(\nabla^j h_{ij}-\partial_i\Tr h)-2(\nabla^j h_{ij}-\partial_i\Tr
h)=0
\lb{14}
\ee
\be
 &&\nabla^k\nabla_i h_{jk}+\nabla^k\nabla_jh_{ik}-\nabla^k\nabla_k
h_{ij}-\nabla_i\nabla_j \Tr h-4h_{ij}\nonumber \\
&&-r^2\partial^2_rh_{ij}+(3-d)r\partial_r
h_{ij}-\gamma_{ij}r^3\partial_r(r^{-2}\Tr h)=0~~.
\lb{15}
\ee
The solution can be taken in the form
\be
h_{ij}(r,x)=\sum_\lambda r^{2-\lambda}\chi^{(\lambda)}_{ij}(x)~~,
\lb{lambda}
\ee
where the sum (or the integral if appropriate) is taken over all appropriate  
$\lambda$.
The analysis is in fact different for $\lambda=0$, $\lambda=1$ and
$\lambda\neq 0\,,\,1$. The equation (\ref{13}) is satisfied automatically
if $\lambda=0$ or $ \lambda=1$ and imposes condition $\Tr \chi^{(\lambda)}=0$ 
when $\lambda \neq 0,\,1$. The equation (\ref{14}) is identically satisfied
if $\lambda=0$ and otherwise imposes condition
\be
\nabla^i\chi_{ij}^{(\lambda)}-\partial_i \Tr \chi^{(\lambda)}=0~,~\lambda\neq 0~~.
\lb{16}
\ee
The third equation, (\ref{15}), takes the form
\be
&&\nabla^k\nabla_i \chi^{(\lambda)}_{jk}+
\nabla^k\nabla_j \chi^{(\lambda)}_{ik}-\nabla^k\nabla_k
\chi^{(\lambda)}_{ij}-\nabla_i\nabla_j \Tr\chi^{(\lambda)} \nonumber \\
&&  +(-2d+m^2_\lambda)
\chi^{(\lambda)}_{ij}+\lambda \gamma_{ij}\Tr\chi^{(\lambda)}=0
\lb{17}
\ee
for any $\lambda$, where the mass term $m_\lambda^2=\lambda (d-\lambda)$ is defined in the same
way as for the scalar field (\ref{4}). 
Notice, that in (\ref{lambda}) all terms can be grouped in pairs,
$r^{2-\lambda}\chi_{ij}^{(\lambda)}$ and 
$r^{2-(d-\lambda)}\chi_{ij}^{(d-\lambda)}$ corresponding to same
mass term $m^2=\lambda(d-\lambda)$. These are two independent solutions to the
second order differential equation.
 
Another case which requires a special treatment is when $m^2={d^2\over 4}$ and
there is only one $\lambda={d\over 2}$ which is related to this mass and hence
only one radial function $r^{2-{d\over 2}}$. Since the second order
differential equations should have two independent solutions there must be
another solution which is not of the form $r^{2-\lambda}$. This second
solution is $r^{2-{d\over 2}}\ln r$ so that in this case we should search 
the solution to the gravitational equations in the form
\be
h_{ij}=r^{2-{d\over 2}}\left(\chi^{(d/2)}_{ij}(x)+\varphi^{(d/2)}_{ij}(x)\ln
  r\right)~~.
\lb{s1}
\ee
Inserting this into equations (\ref{13}), (\ref{14}) and (\ref{15}) 
we find that the fields $\chi^{(d/2)}_{ij}$ and $\varphi^{(d/2)}_{ij}$ decouple
from each other in the gravitational equations and are thus independent
functions.
When $d\neq 2$ both tensors $h^{(d/2)}_{ij}(x)$ and 
$\varphi^{(d/2)}_{ij}(x)$ are transverse and traceless and satisfy equation
(\ref{17}) with $m^2={d^2\over 4}$. When $d=2$ the tensor $\varphi^{(1)}_{ij}$
is still transverse and traceless while the trace of $\chi^{(d/2)}_{ij}$ is not restricted.

Putting things together, we find that depending on $\lambda$ the equations
reduce to one of the following form

\bigskip

\noindent\underline{$\lambda=0$}: \ \ \ \ \ $\nabla^j\chi^{(0)}_{ij}$ and $\Tr \chi^{(0)}$ are
arbitrary;
\be
\noindent \nabla^k\nabla_i \chi^{(0)}_{jk}+
\nabla^k\nabla_j \chi^{(0)}_{ik}-\nabla^k\nabla_k
\chi^{(0)}_{ij}-\nabla_i\nabla_j \Tr\chi^{(0)} 
-2d\chi^{(0)}_{ij}=0
\lb{18}
\ee

\bigskip

\noindent\underline{$\lambda=1$}:  
\be 
\nabla^j\chi_{ij}^{(1)}-\partial_i\Tr\chi^{(1)}=0~~,
\lb{19'}
\ee
\be
\nabla^k\nabla_i \chi^{(1)}_{jk}+
\nabla^k\nabla_j \chi^{(1)}_{ik}-\nabla^k\nabla_k
\chi^{(1)}_{ij}-\nabla_i\nabla_j \Tr\chi^{(1)} 
-(d+1)\chi^{(1)}_{ij}+\gamma_{ij}\Tr \chi^{(1)}=0
\lb{19}
\ee

\bigskip

\noindent\underline{$\lambda\neq 0,\, 1$}:  \ \ \ \ \ 
\be
\nabla^j\chi_{ij}^{(\lambda)}=0~~ {\rm and} ~~\Tr\chi^{(\lambda)}=0~~,
\lb{20'}
\ee
\be
\noindent-\nabla^k\nabla_k
\chi^{(\lambda)}_{ij}
+(2+m^2_\lambda)\chi^{(\lambda)}_{ij}=0~~.
\lb{20}
\ee

\bigskip

\noindent\underline{$\lambda=d/2$}: \ \ \ \ \ the solution takes the
form (\ref{s1}) where depending on value $d$ we have one of the following possibilities

\medskip

\noindent\underline{$d\neq 2$} \ \ \ \ \  
both $\chi^{(d/2)}_{ij}$ and $\varphi^{(d/2)}_{ij}$ satisfy equations
(\ref{20'}) and (\ref{20}) with $m^2=d^2/4$.

\medskip

\noindent\underline{$d=2$} \ \ \ \ \ tensor $\varphi^{(d/2)}_{ij}$ satisfies
equations (\ref{20'}) and (\ref{20}) with $m^2=1$ while the tensor
$\chi^{(d/2)}_{ij}$ satisfies equations (\ref{19'}) and (\ref{19}) and thus should
be identified with $\chi_{ij}^{(1)}$ when $d=2$. 

\bigskip

\noindent These are equations for $\chi^{(\lambda)}_{ij}(x)$ considered as some
symmetric tensor fields on de Sitter spacetime. It is not difficult to
recognize that eq.(\ref{18}) is in fact equation for the massless graviton
on de Sitter space. In particular, it is invariant under the usual gauge
transformations,
\be
\chi^{(0)}_{ij}\rightarrow \chi^{(0)}_{ij}+\nabla_i\xi_j+\nabla_j\xi_i~~.
\lb{*}
\ee
The $\lambda=0$ perturbations  thus describe deformations 
of Einstein space with positive constant curvature. This is of course
consistent with the asymptotic analysis (\ref{9}). The equations (\ref{20})
for linear perturbations characterized by $\lambda\neq 0,\,1$ on the other
hand describe the massive graviton  on $(d+1)$-dimensional de Sitter
space. No gauge symmetry is present in this case.

The case $\lambda=1$ is a bit tricker. Both equations (\ref{19'}) and
(\ref{19}) are invariant under gauge transformations
\be
\chi^{(1)}_{ij}\rightarrow \chi^{(1)}_{ij}+\nabla_i\nabla_j \xi+\gamma_{ij}\xi
\lb{gauge}
\ee
generated by some scalar function $\xi(x)$. This signals that equation (\ref{19})
is some field equation which is already known in the literature.
 Indeed, it is the equation which describes the spin-two 
partially massless field\footnote{I
  thank K. Skenderis for pointing this out to me.}. In the context of adS/CFT
and dS/CFT dualities it was considered in \cite{Witten}, \cite{Deser} and 
\cite{Deser1}. Note that equation (\ref{19'}) arises as a constraint from the field
equation (\ref{19}). It is interesting to note that the gauge transformation 
(\ref{gauge}) has a natural origin from the Minkowski space perspective.
The $(d+2)$-dimensional diffeomorphism preserving the form (\ref{11'})
of Minkowski metric takes the following form, as was found in \cite{dBS},
\be
\xi^r=\alpha(x)~~,~~~\xi^i={1\over r}\gamma^{ij}(x)\partial_j\alpha (x)~~,
\lb{diff}
\ee
where $\alpha(x)$ is arbitrary function, 
the linear perturbations $h_{ij}(r,x)$ of Minkowski metric then changes as follows
\be
\delta_\alpha h_{ij}=2r(\nabla_i\nabla_j\alpha+\gamma_{ij}\alpha)~~.
\lb{hdiff}
\ee
So that this diffeomorphism acts only on the $\lambda=1$ component in
$h_{ij}(r,x)$ the transformation law for which is identical to (\ref{gauge})
(after identifying $\xi=2\alpha$).

Thus, in this first step the gravitational equations on Minkowski space-time
reduce to a set of massless and  massive graviton field equations on de Sitter
space one dimension lower. This is in strict similarity with the scalar field case.
 
\subsection{Second step:  solving field equations  on de Sitter slice}
In the next step we want to solve the equations (\ref{18}), (\ref{19}) and
(\ref{20}) for all values of $\lambda$.  One way to do it is to develop 
an expansion in powers of $e^\tau$ starting from the 
boundary $S^-_d$ on the de Sitter slice. One 
first takes the linear perturbation in the form
$\chi^{(\lambda)}_{ij}(x)=\sum_\kappa 
e^{-(\sigma_{ij}-\kappa)\tau}\chi^{(\lambda, \kappa)}_{ij}(\theta)$, where
$\sigma_{ab}=2$, $\sigma_{\tau a}=1$, $\sigma_{\tau\tau}=-2$  and 
$\chi^{(\lambda,\kappa)}_{ij}(\theta)$ is set of functions on d-sphere, 
and look for certain
values of $\kappa$ for which coefficients $\chi^{(\lambda,\kappa)}_{ij}(\theta)$
are not completely determined by previous terms in the expansion, they then form 
the boundary data to be specified on $S^-_d$. In all cases it is found 
that appropriate values are
$$
\kappa=\lambda ~~~{\tt or}~~~\kappa=d-\lambda~~.
$$
In fact we can do better than just an expansion - we can solve the gravitational
field equations exactly.
The details again depend on the value of $\lambda$\footnote{In the sub-sections
4.2.1 and 4.2.2 we drop the subscript $\lambda$ in the components $\chi_{ij}$.
We hope this should not cause confusion since value of $\lambda$ is 
explicitly indicated in the heading of each sub-section.}.

\subsubsection{$\lambda=0$: Massless graviton in de Sitter space}
The equation for the perturbations in this case is equation for massless
graviton on de Sitter space which has  usual gauge freedom (\ref{*}).
We choose this freedom to further fix the gauge. Specifically we impose
conditions:$\chi_{\tau\tau}=0$ and $\chi_{\tau a}=0$, $a=1,2,.., d$.  So that
the only non-vanishing components are $\chi_{ab}(\tau,\theta)$. Then we have
that $\Tr \chi=e^{-A(\tau)} \tr \chi$, $\tr \chi=\beta^{ab}\chi_{ab}$.
The equations (\ref{18}) then take the form
\be
 &&e^{A}\partial_\tau^2[e^{-A}\tr \chi]-A'^2\tr
\chi+A'\partial_\tau\tr\chi=0
\lb{21}
\ee
\be
&& \partial_\tau [\nabla^b\chi_{ba}-\partial_a\tr \chi]-A'(\tau)
[\nabla^b\chi_{ba}-\partial_a\tr \chi]=0
\lb{22}
\ee
\be
&& e^{-A}[\nabla^c\nabla_a\chi_{bc}+\nabla^c\nabla_b\chi_{ac}-\nabla^c\nabla_c
\chi_{ab}-\nabla_a\nabla_b\tr \chi] \nonumber \\
&&+\partial_\tau^2\chi_{ab}+{d-4\over 2}A'\partial_\tau
\chi_{ab}+\chi_{ab}[A'^2-2d]+{1\over 2}\beta_{ab}(A'\tr h'-A'^2\tr h)=0~~,
\lb{23}
\ee
where covariant derivative $\nabla_a$ is with respect to metric
$\beta_{ab}(\theta)$
on sphere $S^-_d$. Recall that in (\ref{21}), (\ref{22}), (\ref{23}) we have
that
$$
e^{A(\tau)}=\cosh^2\tau~~.
$$
Equations (\ref{21}) and (\ref{22}) are solved immediately and we find that
\be
&&\tr \chi=B(\theta)\cosh\tau\sinh\tau+D(\theta)\cosh^2\tau \nonumber \\
&&\nabla^b\chi_{ab}-\nabla_a\tr \chi=C_a(\theta) e^{A(\tau)}~~,
\lb{24}
\ee
where the integration constants $D(\theta)$ and $B(\theta)$ are some functions on sphere $S^-_d$ and $C_a(\theta)$
is arbitrary vector field on $S^-_d$. As a consequence of (\ref{24}) we have
\be
&&\tr \chi'=A'\tr\chi+B \nonumber \\
&&\tr\chi''=(A''+A'^2)\tr\chi+A'B
\lb{25}
\ee
Taking trace of equation (\ref{23}) with respect to metric
$\beta_{ab}(\theta)$  and using (\ref{24}) and (\ref{25}) we find the relation
between ``integration constants'' $C_a(\theta)$ and $D(\theta)$:
\be
D(\theta)={1\over d-1}\nabla^a  C_a(\theta)~~.
\lb{26}
\ee
The metric $\beta_{ab}$ on sphere $S^-_d$ is maximally symmetric one for which the
curvature tensor reads
\be
&&R_{cabd}=\beta_{cb}\beta_{ad}-\beta_{cd}\beta_{ba} \nonumber \\
&&R_{ab}=(d-1)\beta_{ab}~~.
\lb{1-1}
\ee
Commuting the covariant derivatives with the help of identity
\be
\nabla_c\nabla_a
\chi^c_b-\nabla_a\nabla_c\chi^c_b=d\chi_{ab}-\beta_{ab}\tr\chi
\lb{1-2}
\ee
we find that equation (\ref{23}) (after substituting (\ref{24}))
can be written in the form
\be
\chi''_{ab}+{(d-4)\over \coth\tau}\chi'_{ab}+
{(4-2d)\over \coth^2\tau}\chi_{ab}-{1\over\cosh^2\tau}\nabla^2 \chi_{ab}
-{1\over \coth\tau}\ {\cal B}_{ab}+{\cal F}_{ab}=0~~,
\lb{1-3}
\ee
where we define
\be
&&{\cal B}_{ab}=\beta_{ab} B-\nabla_a \nabla_b B \nonumber \\
&&{\cal F}_{ab}=\nabla_aC_b+\nabla_b C_a+\nabla_a\nabla_b D-2\beta_{ab}D~~,
\lb{1-4}
\ee
and $D$ is defined in (\ref{26}).

As usual the general solution to the inhomogeneous differential equation
(\ref{1-3}) is sum of general solution of homogeneous 
(${\cal B}_{ab}={\cal F}_{ab}=0$) equation and  a particular solution of
the inhomogeneous equation, i.e.
\be
\chi_{ab}=\chi_{ab}^{(\tt hom)}+\chi_{ab}^{(\tt inh)}~~.
\lb{1-5}
\ee
A solution of the inhomogeneous equation can be easily found and it takes the
form
\be
\chi_{ab}^{(\tt inh)}=\chi^{({\cal F})}_{ab}~\cosh^2\tau+\chi^{({\cal B})}_{ab}~
\sinh\tau\cosh\tau~~,
\lb{1-6}
\ee
where
\be
\chi^{({\cal F})}_{ab}={1\over \nabla^2-2} {\cal F}_{ab}~,~~
\chi^{({\cal B})}_{ab}={1\over d-\nabla^2} {\cal B}_{ab}~~.
\lb{1-7}
\ee
Since $\tr {\cal F}=(\nabla^2-2)D$ and $\tr{\cal B}=(d-\nabla^2)B$ we have
that
$$
\tr\chi_{}^{(\tt inh)}=D\cosh^2\tau+B\sinh\tau\cosh\tau
$$
and hence (taking into account (\ref{24})) the homogeneous part in (\ref{1-5})
should be traceless,
$$
\tr\chi_{}^{(\tt hom)}=0~~.
$$
Similarly we can analyze the divergence of (\ref{1-6}). For that we need to
know the commutation relation of covariant derivative $\nabla$ and 
Laplace type operator $\nabla^2=\nabla^a\nabla_a$ 
as acting on
symmetric tensor $\chi_{ab}$. Useful relation for these purposes is the following
\be
\nabla_b\nabla^2 \chi^b_a-\nabla_a\nabla^2\tr\chi= \nabla^2
(\nabla_b\chi^b_a-\nabla_a \tr
\chi)+(d+1)\nabla_b\chi^b_a+(d-3)\nabla_a\tr\chi~~.
\lb{1-8}
\ee
Using this relation (and after some algebra) we find that
(\ref{1-7}) satisfy
\be
&&(1-\nabla^2)(\nabla^b\chi^{(\cal B)}_{ab}-\partial_a\tr\chi^{(\cal
  B)})=0\nonumber \\
&&(\nabla^2+d-1)(\nabla^b\chi^{(\cal F)}_{ab}-\partial_a\tr\chi^{(\cal
  F)})=(\nabla^2+d-1) C_a~~.
\lb{1-9}
\ee
Resolving these equations and ignoring the homogeneous part 
we find that
\be
&&(\nabla^b\chi^{(\cal B)}_{ab}-\partial_a\tr\chi^{(\cal
  B)})=0 \nonumber \\
&&(\nabla^b\chi^{(\cal F)}_{ab}-\partial_a\tr\chi^{(\cal
  F)})=C_a(\theta)~~.
\lb{1-10}
\ee
Thus the homogeneous part in (\ref{1-5}) should be transverse and traceless,
\be
\nabla^b\chi^{(\tt hom)}_{ab}=0~,~~\tr\chi^{(\tt hom)}=0~~.
\lb{1-11}
\ee
Its exact form  can be easily found 
\be
&&\chi^{(\tt hom)}_{ab}=(\cosh\tau)^{5-d\over 2}[
A_0(\nabla^2)P_{d-1\over 2}^{\sqrt{9-2d+d^2-4\nabla^2}\over
  2}(\sqrt{1-\cosh^2\tau})
f_{ab}(\theta) \nonumber \\
&&+B_0(\nabla^2)Q_{d-1\over 2}^{\sqrt{9-2d+d^2-4\nabla^2}\over
  2}(\sqrt{1-\cosh^2\tau})\psi_{ab}(\theta)]
\lb{1-12}
\ee
where $P^\mu_\nu(z)$ and $Q^\mu_\nu(z)$ are Legendre functions and 
$f_{ab}(\theta)$ and $\psi_{ab}(\theta)$ are any transverse-traceless
tensors,
\be
\tr f=\tr \psi=0~,~~\nabla^a f_{ab}=\nabla^a \psi_{ab}=0~~.
\lb{1-13}
\ee
These conditions guarantee that the homogeneous part (\ref{1-12})
of the solution is transverse and traceless. Although the tracelessness is
quite obvious the transverseness should be verified.
Indeed, using the identity (\ref{1-8}) for a traceless tensor
we find that
\be
\nabla^b {\cal P}(\nabla^2)\chi_{ab}={\cal P}(\nabla^2+d+1)\nabla^b\chi_{ab}
\lb{1-14}
\ee
is valid for any function ${\cal P}(\nabla^2)$ of the Laplace operator $\nabla^2$.
Applying this relation to (\ref{1-12}) we find that
condition $\nabla^b\chi_{ab}^{(\tt hom)}=0$ is equivalent to conditions
$\nabla^bf_{ab}=\nabla^b\psi_{ab}=0$.

The eq.(\ref{1-5}) where the inhomogeneous part is given by (\ref{1-6})-(\ref{1-7})
and the homogeneous part has the form (\ref{1-12}) is the general exact
solution to the set of gravitational equations (\ref{21})-(\ref{23}). As it
stands expression (\ref{1-12}) is highly non-local since it contains a very
complicated function of operator $\nabla^2$. However, in the expansion in 
powers of $e^\tau$ (when $\tau\rightarrow -\infty$) few first terms are local.

The ``constants'' $A_0(\nabla^2)$ and $B_0(\nabla^2)$ in (\ref{1-12})
can be chosen in a way that expansion in powers of $e^{\tau}$, or equivalently
in powers of new variable $\rho=4 e^{2\tau}$, takes the form
$$
\chi^{(\tt hom )}_{ab}={1\over \rho}\left(f_{ab}(\theta)+O(\rho)\right)
+{\rho^{d/2}\over \rho}\left( \psi_{ab}(\theta)+O(\rho)\right)~~.
$$
Combining this with  the analogous expansion for the complete solution (\ref{1-5})
$$
\chi_{ab}={1\over \rho}(\chi^{(0)}_{ab}(\theta)+O(\rho))+
{\rho^{d/2}\over \rho}(\chi^{(d)}_{ab}(\theta)+O(\rho))~~,
$$
where $\chi^{(0)}_{ab}$ has the meaning of deformation of the metric on sphere
$S^-_d$ and $\chi^{(d)}_{ab}$ is related to the stress tensor of the 
dual CFT living on the sphere $S^-_d$, we find that
\be
&&\chi^{(0)}_{ab}=f_{ab}(\theta)+{1\over \nabla^2-2}{\cal F}_{ab}+{1\over
  \nabla^2-d}{\cal B}_{ab}~, \nonumber \\
&&\chi^{(d)}_{ab}=\psi_{ab}(\theta)~~.
\lb{1-15}
\ee
Thus the so far undetermined ``integration constants'' $C_a(\theta)$ and
$B(\theta)$ in (\ref{24}) can be related to the trace and divergence of the deformation
$\chi^{(0)}_{ab}$ 
\be
&&C_a=\nabla^b\chi^{(0)}_{ab}-\partial_a\tr\chi^{(0)}~, \nonumber \\
&&B={1\over (d-1)}(\nabla^a\nabla^b\chi^{(0)}_{ab}-\nabla^2\tr \chi^{(0)})-
\tr\chi^{(0)}~~.
\lb{1-16}
\ee
The first equation in (\ref{1-15}) can be viewed as a way to represent arbitrary symmetric
tensor $\chi^{(0)}_{ab}$ in terms of its trace, divergence and the 
transverse-traceless part.

\bigskip

\noindent{\bf d=2 case is special}. In this case there takes place the following

\medskip

\noindent{\it {\tt Statement:} 
If $\chi_{ab}$ is any tensor on two-dimensional manifold such that
$\tr\chi=0$ and $\nabla^b\chi_{ab}=0$ then 
$$
\nabla^2\chi_{ab}=R~\chi_{ab}~~,
$$
where $R$ is the scalar curvature of the manifold.}

\medskip
This statement can be verified by brut-force calculation.
For sphere we have $R=2$ so that any  transverse-traceless tensor in two dimensions
is an eigen-function of Laplace operator $\nabla^2$ with the eigen-value
$2$. This means that we can make a substitution $\nabla^2=2$
everywhere in (\ref{1-12}). The Legendre functions then become trigonometric
functions and the homogeneous part of the solution reads
\be
\chi^{(\tt hom)}_{ab}=\tilde{f}_{ab}\cosh^2\tau+\tilde{\psi}_{ab}\cosh\tau\sinh\tau~~.
\lb{1-17}
\ee
Combining this with the inhomogeneous part (\ref{1-6}) we find that the
total solution (\ref{1-5}) in two dimensions being expressed
in terms of variable $\rho$ has 
only few terms. This is of course consistent with the more general result
obtained in \cite{KS}  that in d+1=3 the solution to
Einstein equations with nonzero cosmological constant has $\rho$-expansion which
terminates on the first three  terms. Here we have proven this for the  
perturbations. The proof however is rather non-trivial since it was not quite
clear from the expression (\ref{1-12}) how the complicated Legendre functions
may reduce to just  few exponential terms even after we put d=2 in (\ref{1-12}).
The above {\tt Statement} was crucial for the demonstration of the consistency. 

Summarizing this subsection, the arbitrary symmetric tensor
$\chi^{(0)}_{ab}(\theta)$ describing deformation of the metric structure on 
d-sphere and the transverse-traceless tensor $\psi_{ab}(\theta)$
related to the stress tensor of the dual CFT are the holographic data
to be specified on d-sphere $S^-_d$ which completely determine 
the $(d+2)$-dimensional Ricci flat metric in the sector $\lambda=0$.

In the holographic pair $(\cab^{(0)},\psi_{ab})$ the function $\cab^{(0)}$
represents a source which on boundary $S^-_d$ couples to a dual operator 
represented by $\psi_{ab}(\theta)$. The coupling then is as follows
\be
\int_{S^-_d}\cab^{(0)}\psi^{ab}~~.
\lb{coup}
\ee
The gauge invariance  (\ref{*}) is usual coordinate invariance on the d-sphere
$$
\delta_\xi\cab^{(0)}=\nabla_a\xi_b+\nabla_b\xi_a~~,
$$
where $\xi$ is a vector on $S^-_d$, under which (\ref{coup}) is supposed to be
invariant.
This imposes constraint $\nabla^a\psi_{ab}=0$ on the dual operator that also
motivates its interpretation as a stress-tensor. This condition is  what
we also get by solving the massless graviton field equation (see (\ref{1-13})).

\subsubsection{$\lambda=1$: Partially massless graviton in de Sitter space}
The  equations for perturbations in this case are collected in Appendix B.
As was discussed above these equations describe a partially massless graviton
field in de Sitter space of dimension d+1. 
This equation has gauge symmetry (\ref{gauge}).
In order to fix the gauge-independent degrees of freedom  
we may want to impose some gauge conditions. A possible condition to impose is 
$$
\Tr\chi=-\chi_{\tau\tau}+e^A\tr\chi=0~~.
$$
It is the gauge suggested in \cite{Witten}. 
Another possible way to impose gauge fixing constraint is to demand that
\be
\ctt=0~~.
\lb{30}
\ee
Looking at the transformation for the component
$\ctt$,
\be
\delta_\xi\ctt=\partial_\tau^2\xi-\xi~~,
\lb{g1}
\ee
we find that condition (\ref{30}) restricts the gauge parameter
$\xi(\theta,\tau)$ to take the form
\be
\xi=\xi_0(\theta)e^{-\tau}+\xi_2(\theta)e^\tau~~.
\lb{g2}
\ee
In this sub-section we prefer to use  condition (\ref{30}) and will see
that field equations are considerably simplified in this gauge. 
As we can see from (\ref{g2}),
the condition (\ref{30}) does not fix the components $\xi^{(0)}$ and $\xi^{(2)}$ 
of the gauge parameter so that there still remains some fiducial gauge
invariance. In fact this invariance is important and plays the role similar to
the asymptotic conformal symmetry in the case $\lambda=0$.
On $(\tau a)$- and $(ab)$-components of the perturbation the gauge transformation with
parameter taking the form (\ref{g2}) acts as follows
\be
&&\delta_\xi\cta={1\over \cosh\tau}(\partial_a\xi_0-\partial_a\xi_2) \nonumber \\
&&\delta_\xi\cab=e^{-\tau}[\nabla_a\nabla_b\xi_0+{1\over 2}\beta_{ab}(\xi_0+\xi_2)]
+e^\tau[\nabla_a\nabla_b\xi_2+{1\over 2}\beta_{ab}(\xi_2+\xi_0)]~~.
\lb{g3}
\ee
In the gauge (\ref{30}) the equations of Appendix C are simplified and can be
solved explicitly. Substituting equation (\ref{A1}) into (\ref{A3}) and
recalling that $e^A=\cosh^2\tau$ we find that $\tr \chi=\beta^{ab}\chi_{ab}$
satisfies a simple differential equation
\be
\tr \chi''-\tr \chi=0~~,
\lb{31}
\ee
the general solution is
\be
\tr \chi=\alpha(\theta)\cosh\tau+\gamma(\theta)\sinh\tau~~,
\lb{32}
\ee
where $\alpha(\theta)$ and $\gamma(\theta)$ are some integration constants.
Substituting this back to equations (\ref{A1}) and (\ref{A2}) we get that
\be
\nabla^a\chi_{a\tau}={\gamma(\theta)\over \cosh\tau}
\lb{33}
\ee
and
\be
(\nabla^b\chi_{ba}-\partial_a\tr\chi)=\cosh^2\tau(\partial_\tau
\cta+d{\sinh\tau\over \cosh\tau}\cta)~~.
\lb{33*}
\ee
Taking one more divergence of eq.(\ref{33*}) we get
\be
\nabla^a\nabla^b\chi_{ab}-\nabla^2\tr\chi=(d-1)\gamma(\theta)\sinh\tau~~.
\lb{34}
\ee
The equations (\ref{32}) and (\ref{34}) tell us that in the expansion of 
the perturbation $\chi_{ab}$ in powers of $e^{\tau}$ all terms, except the first
three terms, are traceless and partially conserved.   
The partial conservation  is very important (see \cite{Witten}) in the theory 
of partially massless graviton field and for  its relation to a conformal field
theory on the boundary. As was discussed in \cite{Witten} the partial
conservation is directly related to the gauge symmetry\footnote{In paper
  \cite{Witten} only the part of transformations which is due to $\xi_0$ 
was considered.} generated by (\ref{g2}). We discuss this point later in the paper. 
Here we just  note that the functions $\alpha (\theta)$ and
$\gamma(\theta)$  transform as 
\be
&&\delta_\xi\alpha(\theta)=(\nabla^2+d)(\xi_0+\xi_2) \nonumber \\
&&\delta_\xi\gamma(\theta)=\nabla^2(\xi_2-\xi_0)~~.
\lb{g4}
\ee
These functions are the only variables which transform non-trivially under the gauge 
transformation (\ref{gauge}).

Next equation to be solved is (\ref{A4}). Substituting the gauge
condition (\ref{30}), equation (\ref{A2}) and explicit expression for $A(\tau)$ we find
that this equation takes a  simpler form
\be
\chi_{\tau a}''+d{\sinh\tau\over \cosh\tau}\chi_{\tau a}'+
[d-1-{(\nabla^2-1)\over \cosh^2\tau} ]\chi_{\tau
  a}+{\partial_a\gamma\over \cosh^3\tau}=0~~.
\lb{35}
\ee
This equation can be solved explicitly and general solution is a sum 
of a particular solution of the
inhomogeneous equation and general solution of the homogeneous equation,
\be
\chi_{\tau a}= \cta^{(\tt inh)}+\cta^{(\tt hom)}~~,
\lb{36}
\ee
where  we find that
\be
 \cta^{(\tt inh)} ={1\over \cosh\tau}{1\over (\nabla^2-d+1)}\partial_a\gamma
\lb{37}
\ee
and
\be
&&\cta^{(\tt hom)}=(\cosh\tau)^{1-d\over 2}[A(\nabla^2)P_{d-3\over
   2}^{\sqrt{5-2d+d^2-4\nabla^2}\over 2}(-i\sinh\tau)J_{a}(\theta) \nonumber \\
&&+B(\nabla^2)Q_{d-3\over
   2}^{\sqrt{5-2d+d^2-4\nabla^2}\over 2}(-i\sinh\tau)I_{a}(\theta)]~~.
\lb{38}
\ee
This solution should satisfy equation (\ref{33}) and thus  we get some
conditions on the so far arbitrary ``constants'' $J_a(\theta)$ and
$I_a(\theta)$. Using identity (\ref{a4}) we show that 
\be
\nabla^a\chi_{a\tau}^{(\tt inh)}={\gamma(\theta)\over \cosh\tau}~~.
\lb{39}
\ee
So that the homogeneous part of the solution should be covariantly conserved,
$\nabla^a\chi_{a\tau}^{(\tt hom)}=0$. Using identity (\ref{a5})  we find that
the latter condition imposes constraints 
\be
\nabla^aJ_a(\theta)=\nabla^aI_a(\theta)=0~~,
 \lb{40}
\ee
i.e. $J_a$ and $I_a$ are arbitrary covariantly conserved vectors on $S^-_d$.
Choosing $A(\nabla^2)$ and $B(\nabla^2)$ appropriately we find that
(\ref{36}) has asymptotic expansion
\be
\cta=2e^\tau(\cta^{(0)}+O(e^\tau))+e^{\tau(d-1)}(I_a(\theta)+O(e^\tau))~~,
\lb{c1}
\ee
where the rest terms in the expansion are completely determined by these two 
terms and we have that
\be
\cta^{(0)}=J_a+{1\over \nabla^2-d+1}\partial_a\gamma~,~~
\nabla^a\cta^{(0)}=\gamma(\theta)
\lb{c2}
\ee
that is a way to present a vector in terms of its divergence
($\gamma$) divergence-free part $(J_a)$. Thus two vectors: arbitrary 
vector $\cta^{(0)}$  and divergence-free vector $I_a$ form 
the first holographic pair at the level $\lambda=1$.

The only equation left is the equation (\ref{A5}) on the components
$\chi_{ab}$ of the perturbation. After all substitutions made this equation reads
\be
\chi_{ab}''+(d-4){\sinh\tau\over \cosh\tau}\cab'+
[3-d-({4-2d+\nabla^2\over \cosh^2\tau})]\cab +\nonumber \\
2{\sinh\tau\over \cosh\tau}[\nabla_a\chi_{\tau a}+\nabla_b\chi_{\tau b}]+
{1\over \cosh\tau}[\nabla_a\nabla_b\alpha-\beta_{ab}\alpha] +
{\sinh\tau\over \cosh^2\tau}[\nabla_a\nabla_b\gamma-2\beta_{ab}\gamma]=0
\lb{41}
\ee
It is again an inhomogeneous equation, the terms staying in the 
second line in (\ref{41}) play the role of the source for the differential
operator staying in the first line.  The solution is again of
the familiar form
\be
\cab=\cab^{(\tt inh)}+\cab^{(\tt hom)}~~,
\lb{42}
\ee
where the homogeneous part takes the form
\be
&&\cab^{(\tt hom)}=(\cosh\tau)^{5-d\over 2}[A_1(\nabla^2)P_{d-3\over
   2}^{\sqrt{9-2d+d^2-4\nabla^2}\over 2}(-i\sinh\tau)k_{ab}(\theta) \nonumber \\
&&+B_1(\nabla^2)Q_{d-3\over
   2}^{\sqrt{9-2d+d^2-4\nabla^2}\over 2}(-i\sinh\tau)p_{ab}(\theta)]~~.
\lb{43}
\ee
Several terms contribute to the inhomogeneous part
\be
\cab^{(\tt inh)}=\cab^{(\alpha)}+\cab^{(\gamma)}+\cab^{(J)}+\cab^{(I)}~~,
\lb{44}
\ee
where 
\be
\cab^{(\alpha)}={1\over \nabla^2-d}(\nabla_a\nabla_b-\beta_{ab})\alpha(\theta)
\cosh\tau
\lb{45}
\ee
\be
&&\cab^{(\gamma)}={2\over \nabla^2-2d+4}({1\over 2}\nabla_a\nabla_b-\beta_{ab}
\nonumber \\
&&+\nabla_a{1\over \nabla^2-d+1}\nabla_b+\nabla_b{1\over \nabla^2-d+1}\nabla_a)\gamma(\theta)\sinh\tau
\lb{46}
\ee
\be
&&\cab^{(J)}={\cal F}^{(J)}(\nabla^2,\tau)(\nabla_a J_b+\nabla_b J_a) 
\nonumber \\
&&\cab^{(I)}={\cal F}^{(J)}(\nabla^2,\tau)(\nabla_a I_b+\nabla_b I_a)~~,
\lb{47}
\ee
and function ${\cal F}^{(J)}(\nabla^2,\tau)$ (${\cal
  F}^{(I)}(\nabla^2,\tau)$)
is a solution to
differential equation
\be
{\cal F}''+(d-4){\sinh\tau\over \cosh\tau}{\cal F}'+(3-d-{(4-2d+\nabla^2)\over
  \cosh^2\tau}){\cal F}
+2{\sinh\tau\over \cosh\tau}\Phi=0
\lb{48}
\ee
with
$$
\Phi^{(J)}(\nabla^2,\tau)=(\cosh\tau)^{1-d\over 2}A(\nabla^2-d-1)P_{d-3\over
   2}^{\sqrt{9+2d+d^2-4\nabla^2}\over 2}(-i\sinh\tau)
$$
and
$$
\Phi^{(I)}(\nabla^2,\tau)=(\cosh\tau)^{1-d\over 2}B(\nabla^2-d-1)Q_{d-3\over
   2}^{\sqrt{9+2d+d^2-4\nabla^2}\over 2}(-i\sinh\tau)~~.
$$
Eq.(\ref{a4}) was used in deriving $\Phi^{(I)}$ and $\Phi^{(J)}$ from (\ref{38}).
We do not have a closed-form expression for functions ${\cal F}^{(J)}$ and
${\cal  F}^{(I)}$ but the expansion is readily available
\be
&&{\cal F}^{(J)}(\nabla^2,\tau)=e^\tau [{1\over 4-d}+O(e^\tau)] \nonumber \\
&&{\cal  F}^{(I)}(\nabla^2,\tau)=e^{\tau(d-1)}[{1\over d}+O(e^\tau)]~~,
\lb{f1}
\ee
where we keep only the leading terms. The dependence on $\nabla^2$ appears in
the subleading terms. Two cases are special: d=2 and d=4.
The expansion then should be modified
\be
&&{\cal F}^{(J,I)}(\nabla^2,\tau)=e^\tau\tau^2 [1+O(e^\tau)]~,~~{\tt d=2}
\nonumber \\
&&{\cal F}^{(J)}(\nabla^2,\tau)=e^\tau\tau [1+O(e^\tau)]~,~~{\tt d=4}~~.
 \lb{f2}
\ee
In terms of variable $\rho$ it involves a logarithm, $\rho\ln\rho$ and
$\rho^{1/2}\ln\rho$ respectively.
As (\ref{f1}) and (\ref{f2}) indicate $\cab^{(J)}$ and $\cab^{(I)}$ contribute
in a way that $J_a$ and $I_a$ show up in the subleading terms of the total
solution (\ref{42}).

Using identities from Appendix A we can now show that 
\be
&&\nabla^a\nabla^b\chi^{(\alpha)}_{ab}-\nabla^2\tr\chi^{(\alpha)}=0 \nonumber \\
&&\nabla^a\nabla^b\cab^{(J)}=\nabla^a\nabla^b\cab^{(I)}=0 \nonumber \\
&&\nabla^a\nabla^b\chi^{(\gamma)}_{ab}-(\nabla^2+d-1)\tr\chi^{(\gamma)}=0~~.
\lb{49}
\ee
It  indicates that the non-conservation in equation
(\ref{34}) is entirely due to the term $\cab^{(\gamma)}$. 
Similarly for the trace we have that
\be
\tr\chi^{(\alpha)}=\alpha(\theta)\cosh\tau~,~~\tr\chi^{(\gamma)}=\gamma(\theta)\sinh\tau~,~~
\tr\chi^{(J)}=\tr\chi^{(I)}=0~~.
\lb{50}
\ee
Combining (\ref{49}) and (\ref{50}) with (\ref{32}) and (\ref{34})
we conclude that the homogeneous part of
the solution should be traceless and partially conserved,
$\tr\chi^{(\tt hom)}=\nabla^a\nabla^b\cab^{(\tt hom)}=0$.  This gives conditions
\be
\tr k=\tr p=0~~{\tt and}~~\nabla^a\nabla^bk_{ab}=\nabla^a\nabla^bp_{ab}=0
\lb{51}
\ee
for the integration constants $k_{ab}(\theta)$ and $p_{ab}(\theta)$ in (\ref{43}).

Choosing appropriately $ A_1(\nabla^2)$ and $B_1(\nabla^2)$ we find that
(\ref{42}) has expansion
\be
\cab={1\over 2}
e^{-\tau}(\cab^{(0)}(\theta)+O(e^\tau))+e^{(d-3)\tau}(p_{ab}(\theta)+O(e^\tau))~~,
\lb{p1}
\ee
where the rest terms in the expansion are determined by these two terms and
by the functions $J_a$ and $I_a$ which appear in the terms starting with $e^\tau$ and
$e^{(d-1)\tau}$ respectively.
The leading term $\cab^{(0)}$ in (\ref{p1}) has the meaning of boundary value
of the perturbation, we have that 
\be
&&\cab^{(0)}={1\over \nabla^2-d}(\nabla_a\nabla_b-\beta_{ab})\alpha+
{2\over \nabla^2-2d+4}({1\over 2}\nabla_a\nabla_b-\beta_{ab} \nonumber \\
&&+\nabla_a{1\over \nabla^2-d+1}\nabla_b+\nabla_b{1\over
  \nabla^2-d+1}\nabla_a)\gamma(\theta)
+k_{ab}~~.
\lb{p2}
\ee
Thus the functions $\alpha$ and $\gamma$ can be related to the trace and 
the partial non-conservation of the tensor $\cab^{(0)}$
\be
&&\alpha(\theta)=2\tr\chi^{(0)}+{1\over
  d-1}[\nabla^a\nabla^b\cab^{(0)}-\nabla^2\tr\chi^{(0)}]
\nonumber \\
&&\gamma(\theta)=-{1\over
  d-1}[\nabla^a\nabla^b\cab^{(0)}-\nabla^2\tr\chi^{(0)}]
\lb{p3}
\ee
so that (\ref{p2}) is just a way to represent any symmetric tensor in terms of
its trace, partial non-conservation and a traceless and partially conserved
part. 

We did not use  yet the equation (\ref{33*}). This equation imposes certain
relations between the so far independent functions $\cab^{(0)}$, $p_{ab}$,
$\cta^{(0)}$ and $I_a$ appearing in the expansions (\ref{p1}) and (\ref{c1}).
Substituting these expansions in the equation (\ref{33*}) and comparing terms
at the same order of $e^\tau$ on both sides we find the 
relations
\be
\nabla^b\cab^{(0)}-\partial_a\tr\chi^{(0)}=(1-d)\cta^{(0)}
\lb{const1}
\ee
and
\be
\nabla^bp_{ab}=-{1\over 4}I_a~~.
\lb{const2}
\ee

Together with (\ref{36}) equations (\ref{42})-(\ref{46}) give us exact and
complete
solution to the partially massless graviton field equations on de Sitter
space.
We are now in the position to determine the holographic data on the boundary
$(S^-_d)$ of de Sitter space.
Additionally to the pair $(\cta^{(0)},~I_a)$ the two functions 
$(\cab^{(0)},~p_{ab})$ form another holographic pair at the level $\lambda=1$.
This data is subject to constraints (\ref{40}), (\ref{51}) and
(\ref{const1}) and (\ref{const2}). Notice that the partial conservation
is a consequence of relations (\ref{const1}) and (\ref{const2}) and of
condition (\ref{40}).
This completes the holographic data at this level.

In the dS/CFT duality in each pair $(\cta^{(0)},~I_a)$ and 
$(\cab^{(0)},~p_{ab})$ the first function should be considered as a source
which couples on the boundary ($S^-_d$) to quantum operator associated with
the second function of the pair. The couplings thus take the form
$$
\int_{S^-_d} \cta^{(0)}I^a~~{\tt and}~~\int_{S^-_d}\cab^{(0)}p^{ab}~~.
$$
The gauge invariance (\ref{g3}) for the source 
$$
\delta_\xi \cta^{(0)}=2\partial_a(\xi_2-\xi_0) 
$$
$$
\delta_\xi \cab^{(0)}=(\nabla_a\nabla_b~\xi_0+{1\over
  2}\beta_{ab}~\xi_0)+{1\over 2}\beta_{ab}~\xi_2
$$
then implies that the dual operators should satisfy certain constraints: 
vector $I_a(\theta)$ should have vanishing divergence and 
tensor $p_{ab}(\theta)$ should be 
traceless and partially conserved. This is exactly what we see from our
solution (see (\ref{40}) and (\ref{51})). Concluding this sub-section we want
to stress that the description dual to the partially massless graviton 
in de Sitter space does not just contain  a tensor operator $p_{ab}(\theta)$ which is
traceless and partially conserved, as it was suggested in
\cite{Witten}. It should contain  also a 
divergence-free vector operator $I_a(\theta)$  related to
the operator $p_{ab}(\theta)$ according to (\ref{const2}).

\subsubsection{$\lambda\neq 0,1$: Massive graviton in de Sitter space}
The process of solving the field equations in this case goes pretty much in 
a similar fashion as before. One of the field equations (or rather
constraints) is that the perturbation should be traceless (see second equation
in (\ref{20'})). This equation allows to express the component $\ctt^\l$
of the perturbation in terms of the trace of components $\cab^\l$ in the
following way
\be
\ctt^{(\lambda)}=e^{-A(\tau)}\tr\chi^{(\lambda)}~~.
\lb{m1}
\ee
The first equation in (\ref{20'}) then gives a pair of equations (where
(\ref{m1}) has been taken into account) 
\be
\nabla^a\cta^{(\lambda)}=\partial_\tau \tr\chi^{(\lambda)}+{(d-1)\over 2}A'
\tr\chi^{(\lambda)}
\lb{d2}
\ee
and
\be
\nabla^b\cba^\l=e^A({d\over 2}A'\cta^\l+\partial_\tau\cta^\l)~~.
\lb{d3}
\ee
The  field equations (\ref{20}) are collected in
Appendix D. Notice that the $(\tau\tau)$ component of (\ref{20}) is an
equation on the trace $\tr\chi^\l$. With the help of (\ref{d2}) this equation
takes the form
\be
\partial_\tau^2\tr\chi^\l+{d\over
  \coth\tau}\partial_\tau\tr\chi^\l+[\lambda(d-\lambda)-
{\nabla^2\over \cosh^2\tau}]\tr\chi^\l=0
\lb{m2}
\ee
and in fact is identical to the scalar field equation (\ref{4}) on de Sitter space
considered in section 2. The solution takes the form similar to (\ref{phi})
\be
&&\tr\chi^\l=(\cosh\tau)^{1-d\over
  2}[A^{(0)}_\lambda (\nabla^2)P_{{(d-1)\over 2}-\lambda}^{\sqrt{(d-1)^2-4\nabla^2}\over
      2}(-i\sinh\tau) f^\l(\theta) \nonumber \\
&&+B^{(0)}_\lambda (\nabla^2)Q_{{(d-1)\over 2}-\lambda}^{\sqrt{(d-1)^2-4\nabla^2}\over
      2}(-i\sinh\tau) g^\l (\theta)]~~,
\lb{m3}
\ee
where $A^{(0)}_\lambda(\nabla^2)$ and $B^{(0)}_\lambda(\nabla^2)$ are chosen in a way that
the asymptotic behavior of (\ref{m3}) take the form
\be
\tr\chi^\l=e^{\tau(d-\lambda)}(f^\l(\theta)+O(e^\tau))+e^{\tau\lambda}
(g^\l (\theta)+O(e^\tau))~~.
\lb{m4}
\ee
Respectively using (\ref{m1}) we find the asymptotic expansion for the
$(\tau\tau)$ components of the perturbation,
\be
\ctt^\l=4e^{2\tau}\left(e^{\tau(d-\lambda)}(f^\l(\theta)+O(e^\tau))+e^{\tau\lambda}
(g^\l (\theta)+O(e^\tau))\right)~~.
\lb{m4*}
\ee
The $(\tau a)$ component (\ref{d5}) of equation (\ref{20}) takes  a similar
form of inhomogeneous differential equation
\be
\partial_\tau^2\cta^\l+{d\over
  \coth\tau}\partial_\tau\cta^\l+[\lambda(d-\lambda)-
{(\nabla^2-1)\over \cosh^2\tau}]\cta^\l+
2{\sinh\tau\over \cosh^3\tau}\partial_a\tr\chi^\l=0~~.
\lb{m5}
\ee
The solution takes the form of sum of two terms
\be
\cta=\cta^{\tt (hom)}+\cta^{(\tt inh)}~~,
\lb{m6}
\ee
where 
\be
&&\cta^{\tt (hom)}=(\cosh\tau)^{1-d\over
  2}[A^{(1)}_\lambda (\nabla^2)P_{{(d-1)\over 2}-\lambda}^{\sqrt{d^2-2d+5-4\nabla^2}\over
      2}(-i\sinh\tau) J^{(\lambda)}_a(\theta) \nonumber \\
&&+B^{(1)}_\lambda (\nabla^2)Q_{{(d-1)\over 2}-\lambda}^{\sqrt{d^2-2d+5-4\nabla^2}\over
      2}(-i\sinh\tau) I^{(\lambda)}_a (\theta)]~~,
\lb{m7}
\ee
and $I^\l_a(\theta)$ and $J^\l_a(\theta)$ are (at this point) 
arbitrary vectors on d-sphere.
The term $\cta^{(\tt inh)}$ in (\ref{m6}) is due to the inhomogeneity in
(\ref{m5}) caused by the term depending on $\tr\chi^\l$. Although a closed-form
expression for $\cta^{(\tt inh)}$ may be possible to find what we really need
is an expansion of the solution in powers of $e^\tau$,
\be
\cta^{(\tt inh)}(\tau,\theta)=4e^{2\tau}\left(e^{\tau(d-\lambda)}(
{\partial_a f^\l(\theta)\over 2+d-2\lambda}+O(e^\tau))
+e^{\tau\lambda}({\partial_a g^\l(\theta)\over 2\lambda+2-d}+O(e^\tau))\right)~~.
\lb{m8}
\ee
These terms are subleading with respect to those coming from the expansion of
(\ref{m7}) so that for the total solution (\ref{m6}) we have that
\be
\cta^\l=e^{\tau(d-\lambda)}(I_a^{(\lambda)}(\theta)+O(e^\tau))+e^{\tau\lambda}
(J_a^\l (\theta)+O(e^\tau))~~.
\lb{m9}
\ee
The equation (\ref{d6}) 
\be
&&\chi_{ab}''+{(d-4)\over \coth\tau}\cab'+
[\lambda(d-\lambda)+4-2d-({4-2d+\nabla^2\over \cosh^2\tau})]\cab \nonumber \\
&&+2{\sinh\tau\over \cosh\tau}[\nabla_a\chi_{\tau a}+\nabla_b\chi_{\tau a}]-
2{\sinh^2\tau\over \cosh^2\tau}\beta_{ab} \ \tr\chi=0
\lb{m10}
\ee
has solution in the form
\be
\cab^\l=\cab^{(\tt hom)}+\cab^{(\tt inh)}~~,
\lb{m11}
\ee
where the homogeneous part is
\be
&&\chi^{(\tt hom)}_{ab}=(\cosh\tau)^{5-d\over 2}[
A^{(2)}_\lambda (\nabla^2)P_{{d-1\over 2}-\lambda}^{\sqrt{9-2d+d^2-4\nabla^2}\over
  2}(-i\sinh\tau)
f^\l_{ab}(\theta) \nonumber \\
&&+B^{(2)}_\lambda (\nabla^2)Q_{{d-1\over 2}-\lambda}^{\sqrt{9-2d+d^2-4\nabla^2}\over
  2}(-i\sinh\tau)\psi^\l_{ab}(\theta)]~~.
\lb{m12}
\ee
Its expansion (when $\tau\rightarrow -\infty$) is
\be
\cab^{(\tt hom)}=
e^{-2\tau}\left(e^{\tau(d-\lambda)}(\psi_{ab}^{(\lambda)}(\theta)+
O(e^\tau))+e^{\tau\lambda}(f_{ab}^\l (\theta)+O(e^\tau))\right)~~.
\lb{m13}
\ee
The expansion of the inhomogeneous part in (\ref{m11})  can be obtained by
inserting the expansions for $\tr\chi^\l$ (\ref{m4}) and $\cta^\l$ (\ref{m9})
into the equation (\ref{m10}) and taking the leading part in the equation.
We then obtain
\be
\cab^{(\tt inh)}={1\over 2\lambda-d+2} H_{ab}^\l\  e^{\lambda\tau}+{1\over
  d+2-2\lambda}G_{ab}^\l \ e^{(d-\lambda)\tau}~~,
\lb{m14}
\ee
where we keep only the leading terms and defined
$$
H_{ab}^\l=\nabla_aI_b^\l+\nabla_bI_a^\l+\beta_{ab}g^\l
$$
$$
G_{ab}^\l=\nabla_aJ_b^\l+\nabla_bJ_a^\l+\beta_{ab}f^\l~~.
$$
Obviously (\ref{m14}) is subleading with
respect to (\ref{m13}). Thus to the leading order the $(ab)$ components of the
perturbation behave as 
\be
\cab^\l=e^{-2\tau}(e^{\tau(d-\lambda)}\psi_{ab}^{(\lambda)}(\theta)+
e^{\tau\lambda}f_{ab}^\l (\theta))~~.
\lb{m13*}
\ee
At this point we see that the complete solution to the set of equations 
(\ref{m1}), (\ref{d4})-(\ref{d6}) is characterized by the  set of tensors
$f_{ab}^\l$, $\psi_{ab}^\l$, $J^\l_a$, $I^\l_a$, $g^\l$ and $f^\l$ defined  
on the sphere $S^-_d$. 
Now we have to take into account the equations (\ref{d2}) and (\ref{d3}). This
will impose certain relations between these functions. 
Indeed, substituting expansions (\ref{m4}), (\ref{m9}) and (\ref{m13}) 
into (\ref{d2}) and (\ref{d3})  and looking at the leading order we get the relations
\be
\nabla^a J_a^\l=(\lambda-d+1)g^\l ~~{\tt and}~~\nabla^aI_a^\l=(1-\lambda)f^\l
\lb{m14*}
\ee
and
\be
\nabla^b f^\l_{ab}={(\lambda-d)\over 4}J^\l_a ~~{\tt
  and}~~\nabla^b\psi_{ab}^\l=-{\lambda\over 4} I^\l_a~~.
\lb{m15}
\ee
More relations come from the consistency condition 
that trace of (\ref{m11}) should coincide
with (\ref{m3}). Comparing the leading terms in the expansions (\ref{m4}) and
(\ref{m13*}) we find that $\psi^\l_{ab}$ and $f^\l_{ab}$ should be traceless,
\be
\tr\psi^\l=\tr f^\l=0~~.
\lb{m16}
\ee
This also means that the whole homogeneous part (\ref{m12}) should be
traceless so that the non-vanishing trace $\tr\chi^\l$ (\ref{m3})
is entirely due to the 
inhomogeneous part in (\ref{m11}).
To the leading order it can be checked directly (using relations (\ref{m14*})
and expansions (\ref{m14}) and (\ref{m4})).

Several remarks are in order. First, it should be noted that all 
holographic data
satisfying the relations (\ref{m14*}) and (\ref{m15}) can be grouped in pairs:
the one which corresponds to $\lambda$ and another one which corresponds 
to $(d-\lambda)$. Interestingly it can be extended to include the cases 
$\lambda=0$ and $\lambda=1$ which are not of the massive graviton case. 
Then, the pair to the massless graviton $\lambda=0$ is the 
massive graviton with $\lambda=d$ and the pair to partially massless graviton
$\lambda=1$ is the massive graviton with $\lambda=d-1$. Of course, these
pairs are just two independent solutions to the second order radial 
differential equations discussed in section 4.1.
Second, a somewhat degenerate case is  $\lambda={d\over 2}$. Then 
the two independent asymptotic solutions considered in this sub-section
should be $e^{\tau{d\over 2}}$ and 
$e^{\tau{d\over 2}}\tau$. We do not consider in detail this case.
Finally, we note that the asymptotic form (\ref{m13*}) multiplied by the radial
function $r^{2-\lambda}$  with complex $\lambda={d\over 2}+i\alpha$
is exactly what one would have expected for the representation of plane
gravitational waves and is similar to the representation (\ref{5}) for the 
plane waves in the case of the scalar field.

\subsection{Summary}
Let us summarize our rather long analysis.
The metric perturbation over Minkowski space has been represented
in the form
$$
h_{ij}(r,\tau,\theta)=\sum_{(\lambda)}r^{2-\lambda}\chi_{ij}^\l(\tau,\theta)+
r^{2-{d\over 2}}\left(\chi^{(d/2)}_{ij}(x)+\varphi^{(d/2)}_{ij}(x)\ln
  r\right)~~,
$$
where sum over $\lambda$ may contain also integral as it happens when
$\lambda={d\over 2}+i\alpha$ with continuous $\alpha$.
Also, the case of $\lambda={d\over 2}$ is special and involves  a logarithm in
the representation for the perturbation and is written in the above expression
explicitly.

\medskip

\noindent\underline{$\lambda=0$}: components $\ctt$ and $\cta$ vanish by the 
gauge
conditions; the asymptotic behavior for $(ab)$ components is
$$
\cab=e^{-2\tau}[(\cab^{(0)}+..)+(e^{(d-1)\tau}\psi_{ab}+..)]~~,
$$
where $\cab^{(0)}(\theta)$ is arbitrary and has the meaning of 
deformation of the metric structure on d-sphere; $\psi_{ab}(\theta)$ should
satisfy conditions
$$
\tr\psi=0~~{\tt and}~~\nabla^b\psi_{ab}=0~~.
$$

\medskip

\noindent\underline{$\lambda=1$}: component $\ctt=0$ by gauge fixing; the
asymptotic behavior of the non-vanishing components 
$$
\cta=2e^\tau(\cta^{(0)}(\theta)+..)+e^{\tau(d-1)}(I_a(\theta)+..)
$$
$$
\cab={1\over 2}e^{-\tau}[(\cab^{(0)}(\theta)+..)+
e^{(d-3)\tau}(p_{ab}(\theta)+..)]~~,
$$
where the following conditions should be satisfied
$$
\nabla^aI_a=0~~{\tt and}~~\tr p=0~,~~\nabla^a\nabla^bp_{ab}=0
$$
as well as relations
$$
\nabla^b\cab^{(0)}-\partial_a\tr\chi^{(0)}=(1-d)\cta^{(0)}~~{\tt and
  }~~\nabla^bp_{ab}=-{1\over 4}I_a~~.
$$

\medskip

\noindent\underline{$\lambda\neq 0,1$}: all components are non-vanishing, 
the asymptotic behavior is as follows
$$
\ctt^\l=4e^{2\tau}[e^{\tau(d-\lambda)}(f^\l(\theta)+..)+e^{\tau\lambda}
(g^\l (\theta)+..)]
$$
$$
\cta^\l=e^{\tau(d-\lambda)}[(I_a^{(\lambda)}(\theta)+..)+e^{\tau\lambda}
(J_a^\l (\theta)+..)]
$$
$$
\cab^\l=e^{-2\tau}[e^{\tau(d-\lambda)}(\psi_{ab}^{(\lambda)}(\theta)+..)+
e^{\tau\lambda}(f_{ab}^\l (\theta)+..)]~~
$$
with constraints and relations
$$
\tr\psi^\l=\tr f^\l=0
$$
$$
\nabla^a J_a^\l=(\lambda-d+1)g^\l~~{\tt and}~~\nabla^aI_a^\l=(1-\lambda)f^\l
$$
$$
\nabla^b f^\l_{ab}={(\lambda-d)\over 4}J^\l_a ~~{\tt
  and}~~\nabla^b\psi_{ab}^\l=-{\lambda\over 4} I^\l_a~~.
$$
\medskip

\noindent\underline{$\lambda={d\over 2}$}: If $d\neq 2$ both functions
$\chi_{ij}^{(d/2)}$ and $\varphi_{ij}^{(d/2)}$ have the same expansion as in
the case $\lambda\neq 0,1$ considered above; in the case $d=2$  the function
$\chi^{(d/2)}_{ij}$ should be identified with the $\lambda=1$ perturbation.

The coefficients in the above expansions (subject to the above mentioned
constraints)
form the holographic data on the sphere $S^-_d$ which
should be  sufficient for the complete 
reconstruction of the (d+2)-dimensional
Ricci-flat metric. The uncovered holographic data should have interpretation
in terms of the conformal field theory living on sphere $S^-_d$ as well
as from the point of view of the (d+2)-dimensional gravitational physics.
As for the CFT interpretation   the tensors $f^\l_{ab}(\theta)$ and
$\psi^\l_{ab}(\theta)$ are naturally interpreted as an infinite set of
the stress-tensors corresponding to the infinite set of the conformal operators on
d-sphere that represent the matter  degrees of freedom. 
That the stress-tensors are not conserved indicates that the
the dual conformal theory couples to a set of sources (represented by
operators $J^\l_a$ and $I^\l_a$). In this case the 
conservation is replaced by the Ward identity (see \cite{deHaro} for some
discussion of this). On the other hand, the operators $J^\l_a$ and $I_a^\l$ are not
conserved as well due to coupling to the operators $g^\l$ and $f^\l$.
It would be interesting to make more precise the relation between these operators
and the infinite set of operators 
(${\cal O}^>_\lambda$ and ${\cal  O}^<_\lambda$)
representing the matter fields. For that one would have to analyze the coupled
gravity-matter system.

It seems natural to suggest that the holographic data should encode 
information about the
mass and rotation of the asymptotically flat gravitational
configuration. In particular, we expect that $I^\l_a$ and $J^\l_a$
should carry information about the angular momentum.
Also, the data should contain information about 
the energy flow coming through the null-infinity. The latter 
seems to be encoded in the data corresponding to $\lambda={d\over 2}+i\alpha$.
The details however need to be further understood.
In the next section we solve a somewhat simpler problem and analyze
where in the holographic data on $S^-_d$ it is stored the information about
mass of the static gravitational  configuration.

\section{Asymptotic form of the black hole metric}
\setcounter{equation}0
The holographic data which we revealed in the previous section should
encounter for all relevant information about the gravitational physics in
asymptotically flat space-time. In particular it should encode the energy 
balance between the mass of the gravitational configuration and the
energy flow coming through the null-infinity. Thus, we expect that the
Bondi mass can be appropriately re-defined in terms of the described
holographic data. This is a problem for future investigation. Here we solve a
simpler problem of encoding the information about the mass of a static
configuration. We want to see which particular 
term in the $\lambda$-representation of the asymptotic  
metric contains information about the mass.
We start with the known metric of (non-rotating) black hole and then bring it
to the form which is more appropriate for our asymptotic analysis.
The standard form of the metric is
\be
ds^2=-g(\rho)dt^2+g^{-1}(\rho)d\rho^2+\rho^2d\omega^2_{S_d}~~,
\lb{5.1}
\ee
where   the  metric function $g(\rho)$
depends on the space-time dimension. When the space-time dimension is d+2=4
the metric (\ref{5.1}) is the Schwarzschild solution,
\be
g(\rho)=1-{2m\over \rho}~~.
\lb{5.2}
\ee
In higher dimensions the metric is known as Myers-Perry metric \cite{MP},
\be
g(\rho)=1-{2m\over \rho^{d-1}}~~.
\lb{5.3}
\ee
Parameter $m$ is the mass in the case d=2 and is related to the mass
when $d>2$. The metric (\ref{5.1}) should be brought to the form
\be
ds^2=dr^2+r^2(-F(r,\tau)d\tau^2+R(r,\tau)d\omega^2_{S_d})
\lb{5.4}
\ee
in terms of new coordinates $(r,\tau)$. When $r$ goes to infinity the
function $F(r,\tau)$ should approach $1$ while the function 
$R(r,\tau)=\cosh^2\tau$ in this limit so that the standard form of the
flat space-time metric is restored. In this limit the relation between coordinates
$(\rho,t)$ and $(r,\tau)$ is  $\rho=r\cosh\tau$, $t=r\sinh\tau$.

That the metric (\ref{5.1}) is non-flat
manifests in modifying these relations by subleading terms. Respectively, the
subleading terms appear in the $r$-expansion of the functions $R(r,\tau)$
and $F(r,\tau)$. It is rather straightforward although quite tedious
to obtain this expansion. Below we present the result.

\medskip

\noindent\underline{$d=2$}:
\be
&&F(r,\tau)=1+{4m\over \cosh^3\tau}{\ln r\over r}+{2\over r}f(\tau) 
\nonumber \\
&&R(r,\tau)=\cosh^2\tau-{2m\over \cosh\tau}{\ln r\over
  r}+{2\over r}a(\tau)\cosh\tau
\lb{5.5}
\ee

\medskip

\noindent\underline{$d>2$}:
\be
&&F(r,\tau)=(1+{f(\tau)\over r})^2-({d\over d-2}){2m\over
  \cosh^{d+1}\tau}{1\over r^{d-1}} \nonumber \\
&&R(r,\tau)=(\cosh\tau+{a(\tau)\over r})^2
+{2m\over (d-2)\cosh^{d-1}\tau}{1\over r^{d-1}}
\lb{5.6}
\ee
We skip the subleading terms in (\ref{5.5}) and (\ref{5.6}).
In both cases the function $f(\tau)$ is arbitrary and $a(\tau)$ is defined as
$$
a(\tau)=\int f(\tau)\sinh\tau d\tau~~.
$$
As we have already seen, the $d=2$ case (Minkowski space-time has physical
dimension 4) is in many respects special. Equation (\ref{5.5}) indicates
another
peculiarity of $d=2$. Quite surprisingly, the $r$-expansion of the
Schwarzschild metric starts with the term with logarithm, $r^{-1}\ln r$.
We see also  that both in (\ref{5.5}) and (\ref{5.6}) the mass makes 
its first appearance at the level  $\lambda=d-1$. To make connection with our
notations let us re-write the expansions (\ref{5.5}) and (\ref{5.6}) in 
the form
\be
&&r^2\left( \varphi_{ij}^{(1)}r^{-1}\ln r +\chi^{(1)}_{ij} r^{-1}\right)~, ~~{\tt
  d=2} \nonumber \\
&&r^2\left(\chi^{(1)}_{ij}r^{-1}+\chi^{(d-1)}_{ij}r^{-(d-1)}\right)~,~~{\tt d>2}
\lb{5.7}
\ee
where
\be
&&\varphi^{(1)}_{\tau\tau}={4m\over
  \cosh^3\tau}~,~~\varphi^{(1)}_{ab}=-{2m\over \cosh\tau}\beta_{ab}(\theta) \nonumber \\
&&\chi^{(d-1)}_{\tau\tau}=-({d\over d-2}){2m\over
  \cosh^{d+1}\tau}~,~~\chi^{(d-1)}_{ab}={2m\over (d-2)\cosh^{d-1}\tau}
\beta_{ab}(\theta)
\lb{5.8}
\ee
and in both cases the components of $\chi^{(1)}_{ij}$  are 
\be
\chi^{(1)}_{\tau\tau}=2f(\tau)~,~~\chi^{(1)}_{ab}=2a(\tau)\cosh\tau \beta_{ab}(\theta)~~.
\lb{5.9}
\ee
Comparing the expressions (\ref{5.8}) with our analysis we see that components
(\ref{5.8}) are what we called the inhomogeneous part (\ref{m14}) of the perturbation
while the homogeneous part vanishes identically in (\ref{5.8}). Notice also
that the $\lambda=d-1$ solution to the radial differential equations
corresponds to mass term $m^2=d-1$. The second independent solution
is the one with $\lambda=1$.  In the case d=2 the two independent solutions
which correspond to the mass term $m^2=1$ are $r\ln r$ and $r$.
We see that in the expansion (\ref{5.6}) there appear both independent solutions
corresponding to $m^2=d-1$: the $\lambda=d-1$ term ($r\ln r$ term in d=2
case) contains information
about the mass of gravitational configuration while the $\lambda=1$ term 
contains an arbitrary function of $\tau$. Freedom in the choice of this
function is apparently a manifestation of the gauge symmetry (\ref{gauge})
appearing exactly at the level $\lambda=1$. 
Comparing (\ref{5.8}) with our analysis  in  section 4.2.3
we can single out the primary element in the holographic data which 
contains the information about the mass. We find that it is the function 
$g^{(\lambda=1)}(\theta)$ which is proportional to the mass $m$ and brings the
dependence on $m$ into all  metric components. 
Generically, $g^{(\lambda)}(\theta)$ (defined in (\ref{m4}))
is a source for the vector $J^\l_a$ via
equation (\ref{m14*}). But in the case $\lambda=1$ the coefficient in front of
the source vanishes and $J^\l_a$ is divergence-free. On the other hand, the
current $J^\l_a$ plays the role of the source for $f^\l_{ab}$
(\ref{m15}) and since $J^\l_a$ is conserved it means that $f^\l_{ab}$ is 
partially conserved. (It is also traceless by (\ref{m16}).)
This latter property singles out the value $\lambda=d-1$. 
Of course,
all $J^\l_a$ and $f^\l_{ab}$ identically vanish in the solution (\ref{5.5})
and (\ref{5.6}). But they would be non-trivial in the general case of dynamical
situation when there is flow of energy coming through the past null-infinity.
It is certainly interesting to analyze how information about the  flow is encoded in the 
holographic data.

\section{Conclusion}
Minkowski space-time can be reconstructed from some data specified on the
boundary of light-cone. This works in a fashion similar to the known holographic
reconstruction of asymptotically anti-de Sitter space from the boundary data.
In the latter case the holographic pair  consists on  metric representing
the conformal class on the boundary  and the  stress-tensor
of the boundary theory. In the present case since infinitely many adS and dS
slices end  at the boundary of the light-cone we should expect that infinitely
many stress-tensors need to be specified there. Also, since the data should
represent the (d+2)-dimensional physics which does not confine to the boundary
only, these stress-tensors are not expected to be conserved. Rather they should
satisfy the  Ward identities as required by the coordinate invariance.
Indeed, what we have found is the  chains of holographic operators that can be represented
as follows
$$
f^\l_{ab}\stackrel{\nabla}{\rightarrow}J^\l_a\stackrel{\nabla}{\rightarrow}
g^\l
$$
and 
$$
\psi^\l_{ab}\stackrel{\nabla}{\rightarrow}I^\l_a\stackrel{\nabla}{\rightarrow}
f^\l~~.
$$
The label $\lambda$ parameterizes the infinite family
of such operators. Another  expectation is that in the dual theory the central charge 
(coming naively from the radius of each (a)dS slice)    should be continuous function
perhaps parameterized by $\lambda$ so that each stress-tensor 
in the infinite family should have its own central charge. 
However, we do not see this in our analysis
since all tensors $f^\l_{ab}$ and $\psi^\l_{ab}$ are traceless.
Of course, these tensors represent only the linearized part of the 
stress-tensors in the full non-linear problem. But that they are traceless
in the linear order may indicate that the central charge is well-defined
in the dual theory independently of $\lambda$. This should be further
investigated. From the gravitational perspective the above sets of operators
represent the bulk gravitational dynamics and should  describe the mass,
angular momentum and the energy flow. The latter can be due to gravitational
waves and is likely to be represented by the holographic operators with complex
$\lambda$. We finish with a remark that our construction may be a nice
starting point for the quantization of asymptotically flat gravitational field
since the set of the holographic data is obviously even-dimensional 
and seems to be well-suited for introduction of the symplectic structure.

\bigskip

{\bf \large Acknowledgments} 
The work on this project took old-fashionedly long
interval of time during which  author's  daughter Alexandra
was born and author  has managed to  move to a new University.  
It is a pleasure to acknowledge the relaxing and
stimulating atmosphere (very much helpful for a technically involved and
long-lasted project) of the group of Slava Mukhanov at LMU, Munchen, where
this project was started. Author thanks Jan de Boer for fruitful collaboration 
on paper \cite{dBS}. The discussions with M. Anderson, J. Barbon, O. Biquard, 
R. Graham, K. Krasnov, R. Mann, R. Myers, I. Sachs,  K. Schleich,
K. Skenderis and L. Smolin were very useful.
The preliminary results were reported at the 73rd Meeting between Physicists 
and Mathematicians on ``(A)dS/CFT correspondence'' (Strasbourg), September
2003. Special thanks to the organizers for organizing this wonderful
meeting and for their kind persistence which played crucial role in  
the actual finishing this work. At different stages this project was supported
by DFG grants SPP 1096 and Schu 1250/3-1.

\appendix{Some useful identities in $dS_d$ and $S_d$}
\setcounter{equation}0
In this Appendix we collect some usefull commutation relations
of covariant derivative and Laplace type operator $\nabla^2=\nabla^a\nabla_a$
on d-dimensional de Sitter space.
These identities are valid in any signature. In the case of Euclidean signature
the space is d-dimensional sphere. Both spaces are maximally symmetric so that
the Riemann curvature can be expressed in terms of metric $\beta_{ab}$,
\be
&&R_{cabd}=\beta_{cb}\beta_{ad}-\beta_{cd}\beta_{ba} \nonumber \\
&&R_{ab}=(d-1)\beta_{ab}~~.
\lb{1*1}
\ee
The commutation of covariant derivatives on such spaces is  significantly 
simplified. The useful relations are 
\be
\nabla_a\nabla^2 \phi=\nabla^2\nabla_a\phi-(d-1)\nabla_a \phi
\lb{a1}
\ee
for  scalar field $\phi$,
\be
\nabla_a\nabla^2 A_b=(\nabla^2-d+1)\nabla_a
A_b+2\beta_{ab}\nabla^cA_c-2\nabla_bA_a
\lb{a2}
\ee
for vector field $A_a$.  Contracting the indices in (\ref{a2}) we find that
\be
\nabla^a \nabla^2 A_a=(\nabla^2+d-1)\nabla^a A_a~~.
\lb{a3}
\ee
Taking the symmetrization of equation (\ref{a2})  and assuming that $\nabla^a
A_a=0$ we get another useful identity for vector
\be
\nabla_a\nabla^2A_b+\nabla_b\nabla^2A_a=(\nabla^2-d-1)(\nabla_aA_b+\nabla_bA_a)~~.
\lb{a4}
\ee
For symmetric tensor $h_{ab}$ we get the  identity
\be
\nabla_b\nabla^2h^b_a=(\nabla^2+d+1)(\nabla_b h^b_a)-2\nabla_a\tr h~~.
\lb{a5}
\ee

\appendix{Legendre functions: differential equation  \\ and asymptotic behavior}
\setcounter{equation}0
Solution to the differential equation
\be
y''(\tau)+{1-2a\over \coth\tau}y'(\tau)-(b-{c\over \cosh^2\tau})y(\tau)=0~~,
\lb{L1}
\ee
where $a$, $b$, $c$ are some constants, 
is a combination of $P$- and $Q$-Legendre functions:
\be
y(\tau)=(\cosh(\tau))^a\left( C_1
P_{-{1\over 2}+\sqrt{b+(a-{1\over 2})^2}}^{\sqrt{a^2+c}}(-i\sinh\tau)+
C_2
Q_{-{1\over 2}+\sqrt{b+(a-{1\over 2})^2}}^{\sqrt{a^2+c}}(-i\sinh\tau)\right)~~,
\lb{L2}
\ee
where $C_1$ and $C_2$ are arbitrary integration constants.

For large $\xi$ the Legendre functions asymptotically behave
as follows \cite{HTF}
\be
&&P^\mu_\nu(\xi)={(2\xi)^\nu\over \sqrt{\pi}}{\Gamma (\nu+{1\over 2})\over
  \Gamma(\nu-\mu+1)}\left(1+O({\ln\xi\over \xi})\right)~,~~{\tt
  Re}\ \nu>-{1\over 2} \nonumber \\
&&Q^\mu_\nu(\xi)={\sqrt{\pi}\over (2\xi)^{\nu+1}}{\Gamma (\nu+\mu+1)\over
  \Gamma(\nu+{3\over 2})}\left(1+O({\ln\xi\over \xi})\right)
\lb{Legandre}
\ee

\appendix{$\lambda=1$ equations on de Sitter space $dS_{d+1}$}
\setcounter{equation}0
Keeping all components of $\chi_{ij}$, i.e. $\chi_{\tau\tau}$, $\chi_{\tau a}$
and $\chi_{ab}$, the equation (\ref{19'}) splits on two equations
\be
{d\over 2}A'\chi_{\tau\tau}=e^{-A}\left(\nabla^a\chi_{a\tau}+{1\over 2}A'\tr\chi
-\partial_\tau\tr\chi\right)
\lb{A1}
\ee
\be
\partial_a\chi_{\tau\tau}-\partial_\tau\chi_{\tau a}-{d\over 2}A'
\chi_{\tau  a}=
e^{-A}(\partial_a \tr\chi - \nabla^b\chi_{ba})~~,
\lb{A2}
\ee
where $\tr\chi=\beta^{ab}\chi_{ab}$ and $\nabla^a$ is with respect to 
metric $\beta_{ab}$ on d-sphere.

The other group of equations comes from (\ref{19}). For $(\tau\tau )$, $(\tau
a)$ and $(ab)$ components of (\ref{19}) we have that
\be
&&{d\over 2}A'\partial_\tau\chi_{\tau\tau}-{d\over 2}(A'^2-2)\ctt \nonumber \\
&&+e^{-A}\left(\tr \chi -{A'^2\over 2}\tr\chi +e^A\partial_\tau^2(e^{-A}\tr
  \chi)-
\nabla^c\nabla_c \ctt +2A'\nabla^b\ctb\right)=0
\lb{A3}
\ee

\be
&&\partial_\tau^2\cta +A'{(d-2)\over 2}\partial_\tau \cta+[{(d+1)\over
  2}(2-A'^2)-{1\over 2}A'']\cta+{3\over 2}A'\partial_a\ctt-\partial_a\partial_\tau\ctt
\nonumber \\
&&+\partial_a\partial_\tau(e^{-A}\tr \chi) 
-{A'\over 2}\partial_a (e^{-A}\tr
\chi)+A'e^{-A}\nabla^b\cab-e^{-A}\nabla^c\nabla_c\cta=0
\lb{A4}
\ee

\be
&&\partial_\tau^2\cab+{(d-4)\over 2}A'\partial_\tau\cab+\cab[(d+1)-{A'^2\over
  2}(d-1)-A'']\nonumber \\
&&+\beta_{ab}e^A[-{A'^2\over 2}\ctt+{A'\over 2}\partial_\tau\ctt+\ctt]\nonumber
\\
&&\beta_{ab}[-{A'\over 2}e^A\partial_\tau (e^{-A}\tr \chi)-\tr\chi]+A'(\nabla_a\ctb+\nabla_b\cta)
\nonumber \\
&&-e^{-A}\nabla^c\nabla_c\cab-\nabla_a\nabla_b\ctt+e^{-A}\nabla_a\nabla_b\tr\chi=0
\lb{A5}
\ee

\appendix{$\lambda\neq 0,1$ equations on de Sitter space $dS_{d+1}$}
\setcounter{equation}0
For components $(\tau\tau)$, $(a\tau)$ and $(ab)$ of equation (\ref{20}) we
find respectively
\be
&&e^A\partial_\tau^2(e^{-A}\tr\chi^\l) -e^{-A}\nabla^2
\tr\chi^\l 
+2A'\nabla^a\cta \nonumber \\
&&+{d\over
  2}A'\partial_\tau\tr\chi^\l+(2+\lambda(d-\lambda)-(d+{1\over
  2})A'^2)\tr\chi^\l=0
\lb{d4}
\ee
\be
&&\partial_\tau^2\cta^\l+({d\over
  2}-1)A'\partial_\tau\cta^\l+(2+\lambda(d-\lambda)-{d+1\over 2}A'^2-{1\over
  2}A'')\cta^\l \nonumber \\
&&-e^{-A}\nabla^2\cta^\l +
A'\partial_a\ctt^\l+A'e^{-A}\nabla^b\cba^\l=0
\lb{d5}
\ee
\be
&&\partial_\tau^2\cab^\l+({d\over 2}-2)A'\partial_\tau\cab^\l-e^{-A}\nabla^2\cab^\l
+(2+\lambda(d-\lambda)-A''-{(d-1)\over 2}A'^2) \nonumber \\
&&+A'(\nabla_a\ctb^\l+\nabla_b\cta^\l)-{1\over 2}A'^2\beta_{ab} \ \tr\chi^\l=0~~.
\lb{d6}
\ee

\end{document}